\documentclass[12pt]{article}
\usepackage{epsf,amsfonts,amssymb,epsfig,amsmath,graphics,subfig}
\usepackage{hyperref,graphicx, subfig}
\renewcommand{\text}[1]{#1}

\newcommand{\be}{\begin{equation}}
\newcommand{\ee}{\end{equation}}
\newcommand{\ben}{\begin{displaymath}}
\newcommand{\een}{\end{displaymath}}
\newcommand{\bea}{\begin{eqnarray}}
\newcommand{\eea}{\end{eqnarray}}
\newcommand{\ba}{\begin{align}}
\newcommand{\ea}{\end{align}}
\newcommand{\nn}{\nonumber \\}
\newcommand{\bi}{\begin{itemize}}
\newcommand{\ei}{\end{itemize}}

\newcommand{\bbR}{{\mathbb{R}}}
\newcommand{\bbZ}{{\mathbb{Z}}}

\begin{document}

\makeatletter
\renewcommand{\theequation}{\thesection.\arabic{equation}}
\@addtoreset{equation}{section}
\makeatother

\baselineskip 18pt

\begin{titlepage}

\vfill

\begin{flushright}
Imperial/TP/2011/JG/01\\
\end{flushright}

\vfill

\begin{center}
   \baselineskip=16pt
   {\Large\bf  Superfluid black branes in $AdS_{4}\times S^{7}$}
  \vskip 1.5cm
      Aristomenis Donos and Jerome P. Gauntlett\\
   \vskip .6cm
      \begin{small}
      \textit{Blackett Laboratory, 
        Imperial College\\ London, SW7 2AZ, U.K.}
        \end{small}\\*[.6cm]

\end{center}

\vfill

\begin{center}
\textbf{Abstract}
\end{center}

\begin{quote}

We consider the $d=3$ $N=8$ SCFT dual to $AdS_4\times S^7$ when held at
finite temperature and chemical potential with respect to a diagonal $U(1)_R\subset SO(8)$ global symmetry and 
construct black brane solutions of $D=11$ supergravity that are associated with the superfluid instability
with the highest known critical temperature. 
We construct a rich array of solutions using different sub-truncations of $SO(8)$ gauged supergravity finding results
that strongly depend on the truncation used. Our constructions include black brane solutions associated with the Gubser-Mitra instability which preserve the $U(1)_R$ symmetry, and these, in turn, can have further superfluid 
instabilities. In addition, we also construct superfluid black branes that at zero
temperature are domain walls that interpolate between the $SO(8)$ $AdS_4$ vacuum in the UV, in an alternative quantisation, and
the supersymmetric $SU(3)\times U(1)$  $AdS_4$ vacuum in the IR.

\end{quote}

\vfill

\end{titlepage}
\setcounter{equation}{0}


\section{Introduction}\label{secone}
Holography provides a powerful 
framework for analysing the different thermodynamic phases of classes of strongly coupled
CFTs. We will be interested in theories that have a non-vanishing chemical potential with respect
to a global $U(1)$ symmetry. 
At very high temperatures the CFT is described by electrically charged AdS black branes of D=10/11 supergravity in which the U(1) symmetry 
is unbroken.
At lower temperatures there can be additional charged black brane solutions which, provided they minimise the
free energy, describe new thermodynamically preferred phases of the system. If these
black branes carry charged hair,  spontaneously breaking the global U(1) symmetry,  the system is in a superfluid phase.

Following earlier work \cite{Gubser:2008px}\cite{Hartnoll:2008vx}\cite{Hartnoll:2008kx}, such holographic superfluid black branes (also called
holographic superconducting black holes)
were constructed in \cite{Gauntlett:2009dn}\cite{Gauntlett:2009bh}
for the infinite class of $d=3$ CFTs that are dual to skew-whiffed $AdS_4\times SE_7$ solutions 
of $D=11$ supergravity, where $SE_7$ is a seven-dimensional Sasaki-Einstein manifold\footnote{Analogous solutions for $d=4$ CFTS dual to
$AdS_5\times SE_5$ solutions of type IIB supergravity were studied in \cite{Gubser:2009qm}\cite{Gubser:2009gp}\cite{Arean:2010wu}. 
We also note that the possibility of $p$-wave holographic superfluids in
$D=5$ gauged supergravity have been explored in \cite{Aprile:2010ge} while holographic superfluids using probe $D$-branes were constructed in 
\cite{Ammon:2008fc}\cite{Ammon:2009fe}\cite{Peeters:2009sr}. }. 
Recall that these CFTs are not supersymmetric\footnote{A discussion of the dual CFTs for the case of $SE_7=S^7/\mathbb{Z}_k$ can be be found in 
\cite{Imaanpur:2010yk}\cite{Forcella:2011pp}.}
except in the special case that $SE_7=S^7$ in which case the CFT is maximally supersymmetric. 
The superfluid black brane solutions of \cite{Gauntlett:2009dn}\cite{Gauntlett:2009bh}
were constructed using a $D=4$ theory of gravity coupled to a gauge field, a charged scalar and a neutral scalar that can be obtained as a consistent KK truncation on an arbitrary $SE_7$ manifold
\cite{Gauntlett:2009zw}\cite{Gauntlett:2009dn}\cite{Gauntlett:2009bh}. 
The consistency of the truncation
means, by definition, that any solution of the D=4 theory can be uplifted on an arbitrary $SE_7$ to obtain an exact solution of $D=11$ supergravity.
The  high temperature black branes, describing the unbroken phase, are embeddings of the 
well known AdS-RN black brane solution. 
The superfluid black branes, which appear below a critical temperature, 
have the interesting feature that at zero temperature they approach a zero entropy state
with an emergent conformal invariance in the far infra-red which is dual to the non-supersymmetric Pope-Warner $AdS_4\times SE_7$ solution \cite{Pope:1984bd}\cite{Pope:1984jj}.
These solutions provide support for the conjecture that finite entropy density at zero temperature, as in the extremal
AdS-RN black brane solution, is always cloaked by phase transitions. Indeed further evidence for this was found in \cite{Gauntlett:2009bh}: it
was found that after
perturbing the CFT by the relevant operator dual to the neutral scalar field there is a dome of superfluid black brane solutions that precisely cloak the finite entropy density at zero temperature of the unbroken phase black branes.

It would be interesting to know whether or not these black brane solutions describe the full phase structure of the dual 
skew-whiffed CFTS at finite temperature and non-vanishing chemical potential. 
In order to establish this one would need to show that there are no other 
thermodynamically relevant black hole solutions. It is certainly a very interesting possibility that
black holes which break spatial translations and rotations play an important role, but for simplicity, we will restrict all of 
our considerations to black holes that preserve these symmetries i.e. black branes.
One can search for new black brane
solutions that are continuously connected to known ones by looking for unstable modes in a linearised
expansion of fields about a known solution. For the case of minimally coupled charged scalars in the AdS-RN black 
brane background this type of analysis was carried out in \cite{Denef:2009tp} 
and the critical temperatures as a function of the mass and charge of the scalar fields
was determined. However, it is certainly possible that some of the linearised scalar modes are not minimally coupled 
(examples have already been
seen in \cite{Donos:2010ax}) and it also possible that there are instabilities associated with higher spin fields. Having identified
the highest temperature instability one is then faced with the task of constructing the back reacted black brane
solutions, which might be a very difficult task unless there is an appropriate consistent KK truncation. One would then
need to repeat the analysis along the new branch looking for further branches of black branes.
 Ideally one would like to establish the nature of all phase transitions tracking them all the way down to the ultimate zero temperature ground state.
Carrying this out in full detail is thus potentially very involved. On the other hand one might hope that for specific examples, where the KK spectrum is well understood, one might be able to make progress if it turns out that only a few KK modes are relevant and furthermore if all of the relevant black branes can be constructed within a consistent KK truncation. A priori one might even entertain the 
idea that the black branes associated with the highest temperature instability 
exist all the way down to zero temperature. 

In this paper we will investigate some of these issues in some more detail in the context of the 
maximally supersymmetric $d=3$ CFT dual to $AdS_4\times S^7$. We will focus on a diagonal or
``Reeb" $U(1)_R\subset SO(8)$ symmetry, corresponding to what was investigated in \cite{Gauntlett:2009dn}\cite{Gauntlett:2009bh}.
The first point is that the black brane solutions constructed in \cite{Gauntlett:2009dn}\cite{Gauntlett:2009bh} 
are {\it not}\footnote{Note that this is in accord with some
observations of \cite{Bobev:2010ib}. 
Namely that the putative ground state at zero temperature
approaches the Pope-Warner $AdS_4\times S^7$ solution in the far IR, but 
in \cite{Bobev:2010ib} this fixed point was shown to be unstable.}, in fact, thermodynamically 
relevant when $SE_7=S^7$. 
Indeed it was already 
pointed out in \cite{Denef:2009tp} that in the linearised fluctuations about the AdS-RN black brane there is a minimally coupled scalar field that gives rise to superfluid black branes with higher critical temperatures than those constructed in
\cite{Gauntlett:2009dn}\cite{Gauntlett:2009bh}. The scalar mode is dual to an operator in the dual CFT with $U(1)_R$ 
charge 1 (suitably normalised) and conformal dimension $\Delta=1$. 
Since all of the operators arising from KK modes on $S^7$ have $\Delta\ge 1$ we can see from figure 1 of \cite{Denef:2009tp}
that, amongst the minimally coupled scalars, this has the highest critical temperature. It was also noted in \cite{Denef:2009tp} that the critical temperature is
higher than that of the Gubser-Mitra instability \cite{Gubser:2000ec}\cite{Gubser:2000mm}. We will show here that there are also neutral scalars with $\Delta=1$ which are not minimally coupled but are in fact stable. It is plausible, therefore, that the mode with $U(1)_R$ charge 1 and $\Delta=1$ identified in
\cite{Denef:2009tp} 
is the one associated with the highest critical temperature for the AdS-RN black brane for the $S^7$ case, but we leave 
a verification of this to future work. 

Instead here we will focus on constructing the corresponding superfluid black brane solutions. The first point to note
is that the relevant $\Delta=1$ modes are contained within $SO(8)$ gauged supergravity. However, it is technically too challenging for us to construct the black brane solutions in this theory directly. So our first task is to find some simpler 
sub-truncations
of $SO(8)$ gauged supergravity that maintain at least one of the relevant charged $\Delta=1$ modes.
In fact we find two different cases. The first case is the truncation that keeps 
the $SU(3)$ invariant sector of the maximal $SO(8)$ gauged supergravity which was
constructed in \cite{Bobev:2010ib} (building on \cite{Warner:1983vz}). It has bosonic
matter content consisting of six scalar fields and two vector fields and it also includes the truncation found in 
\cite{Gauntlett:2009bh} as a sub-truncation. 
The second case starts with the truncation of \cite{Chong:2004ce} which maintains 
$U(1)^4\subset SO(8)$; we will show that this theory has three
further sub-truncations with one or two vector fields plus some charged and neutral scalars, which we shall then study.

Using a numerical shooting technique we construct the back reacted superfluid black branes branching off from the AdS-RN black brane solution corresponding to the $U(1)_R$ charge 1 and $\Delta=1$ mode becoming unstable. Interestingly we will find that the properties
of the black branes depend on the truncation that one is using. For example, in some truncations the superfluid black branes 
exist for temperatures lower than critical temperature, while in others they appear only
to exist for temperatures greater than the critical temperature. These latter solutions thus provide the first string/M-theory 
embeddings of ``exotic hairy black holes" that were constructed in \cite{Buchel:2009ge}, building on \cite{Gubser:2005ih}, and further explored in \cite{Buchel:2009mf}\cite{Buchel:2010wk}.
Another novel feature which we will observe in our general constructions is that the
superfluid black branes can in sprout additional branches of superfluid black branes leading to an elaborate structure.

The sub-truncations of \cite{Chong:2004ce} also admit instabilities involving neutral fields that were first identified
by Gubser and Mitra in \cite{Gubser:2000ec}\cite{Gubser:2000mm}. Here we will construct the corresponding
back reacted black brane solutions. Depending on the truncation
we again find that they can either exist for temperatures less than or greater than the critical temperature. Furthermore, we will
see that these black branes can also grow additional superfluid black brane branches. We will also see that in a certain truncation
there is a first order phase transition from a superfluid branch onto a Gubser-Mitra branch.

An additional motivation for studying the $SU(3)$ invariant truncation of \cite{Bobev:2010ib} 
was to see if it can provide a setting for superfluid black branes that at zero temperature 
become a domain wall which approach a supersymmetric
vacuum in the IR\footnote{Note that in the context of five-dimensional gauged supergravity
supersymmetric domain wall solutions interpolating between two {\it global}
$AdS_5$ spaces with chemical potentials were constructed in \cite{Bobev:2010de}.}. In particular, the truncation contains 
six different $AdS_4$ vacua including an $SU(3)\times U(1)$ invariant vacuum preserving $N=2$
supersymmetry.
We will find that it is indeed possible to construct superfluid black branes that at zero temperature are
domain walls interpolating between the $SO(8)$ $AdS_4$ vacuum and the $SU(3)\times U(1)$ invariant $AdS_4$ vacuum but only in an alternative quantisation of the bulk theory in the UV, corresponding
to a deformation of the maximally supersymmetric $d=3$ SCFT by double trace operators
 \cite{Klebanov:1999tb}.

The plan of the rest of the paper is as follows. In section 2 we discuss the four truncations of $SO(8)$ gauged supergravity
that we shall study in the sequel. In section 3 we analyse the linearised instabilities including the Gubser-Mitra instability.
In section 4 we construct the back reacted black branes for the four different truncations. In section 5 we switch gears
and construct superfluid black branes in an alternative quantisation that approach the $SU(3)\times U(1)$
invariant vacuum at zero temperature and in the far IR. Section 6 contains some final comments and we
have one appendix which discusses some thermodynamics of the black branes.

{\bf Note:} We have been informed that superfluid black brane solutions of
$D=5$ gauged supergravity will be constructed in \cite{Aprile:2011uq}, which will appear at the same time as this paper.

\section{Consistent Truncations of maximal $SO(8)$ gauged supergravity}\label{sectwo}
The KK reduction of $D=11$ supergravity on $S^7$ can be consistently truncated to maximal $D=4$
$SO(8)$ gauged supergravity \cite{deWit:1986iy}. In this section we discuss various truncations of $SO(8)$ gauged supergravity 
that maintain at least a diagonal Reeb $U(1)_R$ gauge field as well as a charged scalar field 
that is dual to an operator with $\Delta=1$ and $U(1)_R$ charge 1 in the dual SCFT. One of the truncations that we 
discuss (see section \ref{4equal}) will have precisely this
field content. The other truncations have an additional abelian gauge field as well as other scalar fields.

We first discuss the consistent truncation to the $SU(3)$ invariant sector of \cite{Bobev:2010ib} that maintains two $U(1)$ gauge fields, 
followed by the truncation of \cite{Chong:2004ce} that maintains four $U(1)$ gauge fields. 
The truncation of \cite{Chong:2004ce} is actually 
not a consistent truncation in the sense that one needs to restrict to a sub-class of solutions to the equations of motion in order
to be able to uplift to a $D=11$ solution, but this will be sufficient for our purposes. We construct three further
truncations of that of \cite{Chong:2004ce}.
The sub-truncations that we consider
in the remainder of the paper are given in \eqref{su3lag}, \eqref{1+3lag}, \eqref{2+2lag} and {\eqref{4lag}.

\subsection{The $SU\left(3\right)$ invariant sector of $SO(8)$ gauged supergravity}\label{su3one}
The consistent truncation of $SO(8)$ gauged supergravity to the
$SU(3)$ invariant sector leads to an $\mathcal{N}=2$ gauged supergravity coupled to a 
vector multiplet and a hypermultiplet \cite{Bobev:2010ib} (for earlier work see \cite{Warner:1983vz}). 
The complex scalar field $z$ in the vector multiplet 
parametrises the special K\"ahler manifold $\frac{SU\left(1,1\right)}{U\left(1\right)\times U\left(1\right)}$ with metric
\begin{equation}
ds^{2}_{SK}=2g_{z\bar{z}}\,dz\,d\bar{z}=\frac{6}{\left(1-\left|z\right|^{2}\right)^{2}}\,dz\,d\bar{z}\, ,
\end{equation}
while the two complex scalar fields $\zeta_i$ in the hypermultiplet parametrise the quaternioninc K\"ahler manifold
$\frac{SU\left(2,1\right)}{SU\left(2\right)\times U\left(1\right)}$ 
with metric
\begin{equation}
ds^{2}_{QK}=2g_{\zeta_{i}\bar{\zeta}_{j}}d\zeta_{i}\,d\bar{\zeta}_{j}=2\left[\frac{d\zeta_{1}\,d\bar{\zeta}_{1}+d\zeta_{2}\,d\bar{\zeta}_{2}}{1-\zeta_{1}\bar{\zeta}_{1}-\zeta_{2}\bar{\zeta}_{2}}+\frac{\left(\zeta_{1}\,d\bar{\zeta}_{1}+\zeta_{2}\,d\bar{\zeta}_{2} \right)\,\left(\bar{\zeta}_{1}\,d\zeta_{1}+\bar{\zeta}_{2}\,d\zeta_{2} \right)}{\left(1-\zeta_{1}\bar{\zeta}_{1}-\zeta_{2}\bar{\zeta}_{2}\right)^{2}}\right]\, .
\end{equation}
The gauging is determined by the two Killing vectors
\begin{align}
K_{0}&=i \left(\zeta_{1}\partial_{\zeta_{1}}-\zeta_{2}\partial_{\zeta_{2}}\right)+ \mathrm{c.c.}\nn
K_{1}&=i \sqrt{3}\left(\zeta_{1}\partial_{\zeta_{1}}+\zeta_{2}\partial_{\zeta_{2}}\right)+ \mathrm{c.c.}\, .
\end{align}

The bosonic Lagrangian of the $N=2$ supergravity theory can be written
\begin{align}\label{eq:action}
\mathcal{L}&=\mathcal{L}_{EH}+\mathcal{L}_{gauge}+\mathcal{L}_{kin}-e \mathcal{P}\, ,
\end{align}
where the Einstein-Hilbert term is $\mathcal{L}_{EH}=\frac{1}{2}eR$.
For the kinetic and topological terms for the two gauge fields we have
\begin{align}
\mathcal{L}_{gauge}&=-\frac{1}{2}\,\left(\mathrm{Re}\left(\mathcal{N}_{\alpha\beta} \right)\tilde{F}^{\alpha}\wedge \ast\tilde{F}^{\beta}+ \mathrm{Im}\left(\mathcal{N}_{\alpha\beta} \right)\tilde{F}^{\alpha}\wedge \tilde{F}^{\beta}\right)\, ,
\end{align}
where $\tilde{F}^{\alpha}=dB^{\alpha}$,  $\alpha=0,1$ and the $2\times 2$ matrix $\mathcal{N}$ is defined by
\begin{align}
\mathcal{N}_{00}&= \frac{1+2\bar{z}+3z\bar{z}+3z^{2}+2z^{3}+z^{3}\bar{z}}{\left(1+z\right)^{2}\left(1-2z+2\bar{z}-z\bar{z} \right)}\nn
\mathcal{N}_{11}&= \frac{1-2\bar{z}+3z\bar{z}+3z^{2}-2z^{3}+z^{3}\bar{z}}{\left(1+z\right)^{2}\left(1-2z+2\bar{z}-z\bar{z} \right)}\nn
\mathcal{N}_{10}=\mathcal{N}_{01}&= -\frac{2\sqrt{3}z\left(1+z\bar{z}\right)}{\left(1+z\right)^{2}\left(1-2z+2\bar{z}-z\bar{z} \right)}\, .
\end{align}
The kinetic term for the scalars is given by
\begin{equation}\label{thestart}
e^{-1}\mathcal{L}_{kin}=-g_{z\bar{z}}\,\partial_{\mu}z\,\partial^{\mu}\bar{z}-g_{\zeta_{i}\bar{\zeta}_{j}}\,D_{\mu}\zeta_{i}\,D^{\mu}\bar{\zeta}_{j}\, ,
\end{equation}
where $D_{\mu}\zeta_{i}=\partial_{\mu}\zeta_{i}+B^{\alpha}_{\mu}\,K_{\alpha}^{\zeta_{i}}$. Explicitly we have
\begin{align}
D_{\mu}\zeta_{1}&=\partial_{\mu}\zeta_{1}+i\,\left(B^{0}_{\mu}+\sqrt{3}\,B^{1}_{\mu} \right)\,\zeta_{1}\nn
D_{\mu}\zeta_{2}&=\partial_{\mu}\zeta_{2}+i\left(-B^{0}_{\mu}+\sqrt{3}\,B^{1}_{\mu} \right)\,\zeta_{2}\, .
\end{align}
Finally, for the potential $\mathcal{P}$ we define
\begin{align}
\mathcal{W_{+}}&=\left(1-\left|z\right|^{2}\right)^{-3/2} \left(1-\left|\zeta_{12}\right|^{2} \right)^{-2}\,\left[\left(1+z^{3}\right)\left(1+\zeta_{12}^{4}\right)+6z\,\zeta_{12}^{2}\,\left(1+z\right) \right]\, ,
\end{align}
where
\begin{align}
\zeta_{12}=\frac{|\zeta_1|+i|\zeta_2|}{1+\sqrt{1-|\zeta_1|^2-|\zeta_2|^2} }\, ,
\end{align}
after which
\begin{equation}
\mathcal{P}=2\,\left[\frac{4}{3}\left(1-\left|z\right|^{2}\right)^{2}\,\left| \partial_{z}\left|\mathcal{W_{+}} \right|\right|^{2} +\left(1-\left|\zeta_{12}\right|^{2} \right)^{2}\left| \partial_{\zeta_{12}}\left|\mathcal{W_{+}} \right|\right|^{2} -3\left|\mathcal{W_{+}} \right|^{2}\right]\, .
\end{equation}

It will be convenient to introduce the following combination of gauge fields
\begin{align}
B^{0}=\frac{1}{2}\,\left(\sqrt{3}\,A^{0}-A^{1} \right),\qquad
B^{1}=\frac{1}{2}\,\left(A^{0}+\sqrt{3}\,A^{1}\right)
\end{align}
and we will write the corresponding field strengths as  $F^{\alpha}=dA^{\alpha}$. 
(Note that our gauge fields $B ^\alpha$ are denoted as $A^\alpha$ in \cite{Bobev:2010ib}. We have also set their $g$ to one).
As we will discuss, $A^1$ will be the $U(1)_R$ gauge field that will be of most interest.
We will write the complex scalar $z$ in terms of real variables via 
\begin{align}
z=\lambda+i \eta
\end{align}
and we note that $\lambda^{2}+\eta^{2}<1$ and $|\zeta_{1}|^{2}+|\zeta_{2}|^{2}<1$.

\subsubsection{$AdS_4$ vacuua}
This consistent truncation has six different $AdS_4$ vacua, which can be labelled by the residual amount of
the $SO(8)$ symmetry that they preserve. Writing the $AdS_4$ metric in the form
\begin{equation}\label{coordsforads4}
ds_{4}^{2}=-2cr^{2}\,dt^{2}+\frac{dr^{2}}{2cr^{2}}+r^{2}\,\left(dx_{1}^{2}+dx_{2}^{2}\right),
\end{equation}
with $R^2_{AdS}=1/2c$,
the vacua are given by 
\begin{itemize}\label{adsfps}
\item $SO\left(8\right)$\\
$\eta=\lambda=\zeta_1=\zeta_2=0,\quad c=1$
\item $SO\left(7\right)^{+}$\\
$\lambda=\frac{5^{1/4}-1}{5^{1/4}+1},\quad \zeta_1=\frac{1}{2}\left(3-\sqrt{5}\right),\quad \eta =\zeta_2=0,\quad c=5^{3/4}/3$
\item $SO\left(7\right)^{-}$\\
$\eta =-\left(2-\sqrt{5}\right),\quad \zeta_2=1/\sqrt{5},\quad \lambda=\zeta_1=0,\quad c=25\sqrt{5}/48$
\item $G_{2}$\\
$\zeta_1=2-\sqrt{3},\quad \zeta_2=\left(3+2\sqrt{3} \right)^{-1/2}$\\
$\lambda=\frac{1}{4}\,\left(3+\sqrt{3}-3^{1/4}\sqrt{10} \right),\quad \eta =-\frac{1}{4} \frac{\sqrt{2+\sqrt{3}}}{3^{1/4}}\,\left(3+\sqrt{3}-3^{1/4}\sqrt{10} \right)$, \quad$c=\frac{36}{25} \sqrt{\frac{2}{5}\sqrt{3}}$
\item $SU\left(3\right)\times U\left(1\right)$\\
$\lambda=2-\sqrt{3},\quad,\quad \zeta_2=\frac{1}{\sqrt{3}},\quad \eta =\zeta_1=0$,\quad$c=\frac{3\sqrt{3}}{4}$
\item $SU\left(4\right)^{-}$\\
$\lambda= \eta =\zeta_1=0,\quad \zeta_2=\frac{1}{\sqrt{2}}$,\quad $c=\frac{4}{3}$
\end{itemize}
The $SO(8)$ invariant vacuum uplifts to the 
maximally supersymmetric $AdS_4\times S^7$ solution of $D=11$. Note that the $SO(7)^\pm$ vacua and 
the $SU(4)^-$ vacuum are known to be unstable within $SO(8)$ gauged supergravity \cite{deWit:1983gs}\cite{Bobev:2010ib}.
On the other hand the $G_2$ and $SU(3)\times U(1)$ vacua are supersymmetric (being dual to $d=3$ CFTs preserving $N=1$ and $N=2$ supersymmetry, respectively) and are stable. The principal focus of this paper will be the $SO(8)$ vacuum
but the 
$SU(4)^-$ and the 
$SU(3)\times U(1)$ vacua will also appear.

Expanding about the $SO(8)$ invariant vacuum we find that
the masses of the scalar fields are all given by $M^{2}=-4$.
Depending on the boundary conditions that are imposed, these can be
dual to operators with scaling dimensions given by
either $\Delta=1$ or $\Delta=2$. Our focus (apart from in section 5) is on the maximally supersymmetric CFT and the appropriate quantisation has
\begin{align}
\Delta({\cal O}_{\zeta_1})=1,\qquad
\Delta({\cal O}_{\zeta_2})=2,\qquad
\Delta({\cal O}_{\lambda})=1,\qquad
\Delta({\cal O}_{\eta})=2\, ,
\end{align}
with $\zeta_1$ having $U(1)_R$ charge 1, $\zeta_2$ having $U(1)_R$ charge 2 and
$\lambda,\eta$ having $U(1)_R$ charge 0.
This can be seen by tracing the origin of the modes from maximal gauge supergravity. The latter theory has pseudo-scalars transforming
in the ${\bf 35_-}$ (i.e. ${\bf 35_c}$) of $SO(8)$ which are dual to operators with $\Delta=2$ and scalars transforming in the ${\bf 35_v}$
of $SO(8)$ which are dual to operators with $\Delta=1$. Decomposing these representations under $SU(4)^-\times U(1)_R\subset SO(8)$
(under which ${\bf 8_+}\to {\bf 4^-}+{\bf\bar 4^+}$ and
${\bf 8_-}\to {\bf 1^{2+}}+{\bf 1^{2-}}+{\bf 6^0}$) 
and then further decomposing under $SU(3)\times U(1)\subset SU(4)^-$ we find that the pseudo-scalars with $\Delta=2$ give rise to three $SU(3)$ singlets with $U(1)_R\times U(1)$ charges $(\pm 2, 0)$ and $(0,0)$. Similarly the scalars with $\Delta=1$ give rise to three $SU(3)$ singlets
with $U(1)_R\times U(1)$ charges $(\pm 1, \pm 1)$ and $(0,0)$. Next considering eq. (2.38) of \cite{Bobev:2010ib} we see that our gauge field $A^1$ corresponds
to the $SO(8)$ generator proportional to $Diag(i\sigma_2,i\sigma_2,i\sigma_2,-i\sigma_2)$, where $\sigma_2$ is the second Pauli matrix, 
while $A^0$ corresponds to 
$Diag(i\sigma_2,i\sigma_2,i\sigma_2,3i\sigma_2)$. We thus identify $A^1$ with what we call 
the Reeb $U(1)_R$ and $A^0$ (up to a normalisation factor) 
with the other $U(1)$. 
Now, with respect to gauge fields $A^1,A^0$ we have that $\zeta_1$ has charges $(\pm 1,\pm\sqrt{3})$ and $\zeta_2$ has charges $(\pm 2,0)$. We thus conclude that $\zeta_2$
has $U(1)_R$ charge 2, arises from the pseudo scalar sector and should be quantised so that $\Delta=2$. Similarly $\zeta_1$ has $U(1)_R$
charge 1, arises from the scalar sector and should be quantised so that $\Delta=1$. In fact this precisely agrees with the discussion in 
\cite{Bobev:2009ms} (see also footnote 4 of \cite{Bobev:2010ib}) which also shows that we should quantise the neutral scalar $\lambda$, the real part of $z$, so that $\Delta=1$ and
the neutral scalar $\eta$, the imaginary part of $z$, so that $\Delta=2$.

\subsubsection{Additional truncations}
We first observe that we can consistently set $A^0=\zeta_1=\zeta_2=\lambda=\eta=0$ in
the $SU(3)$ invariant Lagrangian \eqref{eq:action} to obtain the
Einstein-Maxwell theory
\begin{equation}\label{emlag}
e^{-1}\,\mathcal{L}=\frac{1}{2}R+6-\frac{1}{4} F^{1}_{\mu\nu}F^{1}{}^{\mu\nu}\, .
\end{equation}
We will be interested in instabilities of the electrically charged AdS-RN black brane solutions of this theory.

Returning to the $SU(3)$ invariant Lagrangian \eqref{eq:action} it is not difficult to see that 
it is possible to consistently truncate each of the charged scalars $\zeta_1,\zeta_2$ or both to zero.
The work of \cite{Gauntlett:2009dn}\cite{Gauntlett:2009bh} constructed superfluid black branes with $\zeta_1=0$ and
$\zeta_2\ne 0$ (dual to operators with $\Delta=2$ acquiring a vev). This was carried out within a consistent
truncation found in \cite{Gauntlett:2009dn}\cite{Gauntlett:2009bh} (building on \cite{Gauntlett:2009zw}) obtained 
from a consistent truncation of $D=11$ on an arbitrary $SE_7$ space involving the breathing mode multiplet.
It was shown in  \cite{Bobev:2010ib}, that the same $D=4$ theory can also be obtained 
as a further consistent truncation of the full $SU(3)$ invariant Lagrangian \eqref{eq:action}
to the $SU(4)^-$ invariant sector obtained by
taking $z$ to be purely imaginary, $z=i\eta$ (i.e. $\lambda=0$), and also setting $\zeta_1=A^0=0$
i.e. ${\cal L}={\cal L}(A^1,\zeta_2,\eta)$.
The resulting equations of motion can be derived from an action\footnote{To
compare with  \cite{Gauntlett:2009bh} one should make the identifications: $A^1= -2 A^{GSW}$, $\zeta_2=(\sqrt{3}/2)\chi^{GSW}$,
$h^{GSW}=-2\eta/(1+\eta^2)$, rescale the metric $g_{\mu\nu}=2 g^{GSW}_{\mu\nu}$ and set $16\pi G=1$.} 
that was given explicitly in \cite{Bobev:2010ib}. One can also additionally truncate to $\eta=0$ and study 
the simpler theory
${\cal L}={\cal L}(A^1,\zeta_2)$ first constructed and studied 
in \cite{Gauntlett:2009dn} (and we recall from \cite{Gauntlett:2009dn} that in order to uplift to D=11 one must impose $F^1\wedge F^1=0$ by hand for this truncation.)

In this paper our main focus will be on the instabilities of the AdS-RN black branes with $\zeta_1\ne 0$ (dual to operators with $\Delta=1$ acquiring a vev). 
In general this field sources the real part of $z$, $\lambda$ and also $A^0$, so we cannot set them to zero. However, to simplify things, we will
set, mostly, the imaginary part of $z$ to zero, $\eta=0$. Note that this is {\it not} in general a consistent truncation at the level of the full equations of motion, however it
is consistent with the equations of motion for the purely electrical and radial ansatz  that we consider in
constructing our black brane solutions 
- see the beginning of section 4. The equations of motion for our ansatz can also be derived by 
substituting the ansatz directly into the Lagrangian \eqref{eq:action} with $\eta=0$:
\begin{align}\label{su3lag}
{\cal L}={\cal L}(g_{\mu\nu},A^0_\mu,A^1_\mu,\zeta_1, \zeta_2,\lambda)\, .
\end{align}
It is consistent to set either of the two charged fields $\zeta_1$, $\zeta_2$ to zero and we will sometimes do that.

\subsection{The truncation of \cite{Chong:2004ce}}\label{popescases}
We now consider a different truncation of $N=8$ $SO(8)$ gauged supergravity found in \cite{Chong:2004ce}.
The action of \cite{Chong:2004ce} maintains four gauge fields, $C^i$, parametrising $U(1)^4\subset SO(8)$, four scalar fields $\varphi_i$ each
charged under a single $U(1)$ and three neutral scalar fields $\phi_a$. 
The action of \cite{Chong:2004ce} (after setting their $2g^2=1$ and relabelling gauge fields) is given by
\begin{align}\label{popelag}
e^{-1}\mathcal{L} & =\frac{1}{2}R-\frac{1}{4}\sum_{i=1}^{4}\left(\partial\varphi_{i}\right)^{2}-\frac{1}{4}\sum_{a=1}^{3}\left(\partial\phi_{a}\right)^{2}-\frac{1}{8}\sum_{i=1}^{4}X_{i}^{-2}\,\left(H^{i}\right)_{\mu\nu}\left(H^{i}\right)^{\mu\nu}\nn
&-\frac{1}{2}\sum_{i=1}^{4}\sinh^{2}\varphi_{i}\left(C^{i}\right)_{\mu}\left(C^{i}\right)^{\mu}-V\, ,
\end{align}
where
\begin{align}
V & =\frac{1}{2}\,\left[\sum_{i=1}^{4}X_{i}^{2}\sinh^{2}\varphi_{i}-\sum_{i\neq j}X_{i}X_{j}\cosh\varphi_{i}\cosh\varphi_{j}\right]\nn
X_{1} & =e^{\frac{1}{2}\left(-\phi_{1}-\phi_{2}-\phi_{3}\right)},\quad
X_{2}  =e^{\frac{1}{2}\left(-\phi_{1}+\phi_{2}+\phi_{3}\right)},\quad
X_{3}  =e^{\frac{1}{2}\left(\phi_{1}-\phi_{2}+\phi_{3}\right)},\quad
X_{4}  =e^{\frac{1}{2}\left(\phi_{1}+\phi_{2}-\phi_{3}\right)}\nn
H^{i} & =dC^{i}\, .
\end{align}
Observe that the equations of motion are invariant under flipping the sign of each of the gauge fields.
This KK truncation is is not fully consistent. However, any solution of the $D=4$ equations of motion that in addition satisfies 
$H^i\wedge H^i=0$ for each $i$, can be uplifted on an $S^7$ using the formulae given in \cite{Chong:2004ce} to obtain a solution of
$D=11$ supergravity. 

The $AdS_4$ vacuum with all scalars set to zero uplifts to the maximally supersymmetric $AdS_4\times S^7$ solution.
The formulae given in \cite{Chong:2004ce} show that the seven scalars are all part of the ${\bf 35_v}$ scalars of
the $SO(8)$ gauged supergravity and in particular should be quantised so that they are dual to operators in the maximal SCFT 
with $\Delta=1$.

We next note that in \eqref{popelag} we can consistently set the scalars to zero $\phi_a=\varphi_i=0$, and maintain a single 
$U(1)_R$ gauge field via\footnote{The opposite sign chosen for $C^4$ is simply to have the same $U(1)$ gauge field
$A^1$ considered in the $SU(3)$ invariant truncation of $SO(8)$ gauged supergravity.} 
\begin{align}
C^{1}=C^{2}=C^{3}=-C^{4} & =\frac{1}{\sqrt{2}}A^{1}\, ,
\end{align}
to obtain the Einstein Maxwell theory \eqref{emlag}. 
We now present some enlarged truncations of \eqref{popelag} that maintain at most two charged scalar fields and 
in addition at most one other vector field.

\subsubsection{1+3 equal charged scalars}
We first consider a sub truncation of \eqref{popelag}  that keeps two charged scalar fields, two gauge fields and a single neutral scalar:
\begin{align}
\phi_{1}&=\phi_{2}=\phi_{3}  =-\phi\nn
\varphi_{1} & =\rho,\qquad\varphi_{2}=\varphi_{3}=\varphi_{4}  =\chi\nn
C^{1} & =\frac{1}{\sqrt{2}}\,\left(A^{1}+\sqrt{3}\, A^{0}\right),\qquad
C^{2}=C^{3}=-C^{4}  =\frac{1}{\sqrt{2}}\,\left(A^{1}-\frac{1}{\sqrt{3}}\, A^{0}\right)\, .
\end{align}
Indeed we find that we can derive the equations of motion for the new degrees of freedom
from the action obtained from the Lagrangian ${\cal L}(g_{\mu\nu},A^0_\mu, A^1_\mu,\rho,\chi,\phi)$
given by
\begin{align}\label{1+3lag}
e^{-1}\mathcal{L}&=  \frac{1}{2}R-\frac{3}{4}\left(\partial\phi\right)^{2}-\frac{1}{4}\left(\partial\rho\right)^{2}-\frac{3}{4}\left(\partial\chi\right)^{2}-\frac{1}{4}\sinh^{2}\rho\,\left(A^{1}+\sqrt{3}\, A^{0}\right)^{2}\nn
&-\frac{3}{4}\sinh^{2}\chi\,\left(A^{1}-\frac{1}{\sqrt{3}}\, A^{0}\right)^{2}
  -\frac{1}{4}e^{-\phi}\cosh\left(2\phi\right)\,\left[\left(F^{0}\right)_{\mu\nu}\left(F^{0}\right)^{\mu\nu}+\left(F^{1}\right)_{\mu\nu}\left(F^{1}\right)^{\mu\nu}\right]\nn
 & -\frac{1}{8}e^{-\phi}\sinh\left(2\phi\right)\,\left[-\left(F^{0}\right)_{\mu\nu}\left(F^{0}\right)^{\mu\nu}-2\sqrt{3}\left(F^{1}\right)_{\mu\nu}\left(F^{0}\right)^{\mu\nu}+\left(F^{1}\right)_{\mu\nu}\left(F^{1}\right)^{\mu\nu}\right]-V
 \end{align}
 with
 \begin{align}
V= & \frac{1}{2}e^{-\phi}\,\left[-6\,\cosh^{2}\chi-6e^{2\phi}\cosh\rho\,\cosh\chi+3\,\sinh^{2}\chi+e^{4\phi}\sinh^{2}\rho\right]
\end{align}
and $F^\alpha=dA^\alpha$. As noted above the maximally supersymmetric quantisation has
\begin{align}
\Delta({\cal O}_{\rho})=
\Delta({\cal O}_{\chi})=
\Delta({\cal O}_{\phi})=1\, ,
\end{align}
with $\rho,\chi$ both having $U(1)_R$ charge 1 and $\phi$ having $U(1)_R$ charge 0.

We can make contact with part of the $SU(3)$ invariant truncation of the last subsection. 
Specifically it is consistent to further set $\chi=0$ and after redefining 
\begin{align}
\lambda  =\tanh\left(\frac{\phi}{2}\right),\qquad
\zeta_{1}  =\tanh\left(\frac{\rho}{2}\right)\, ,
\end{align}
we recover a truncation of the Lagrangian 
\eqref{su3lag} after setting $\zeta_2=0$ and $\zeta_1$ real.
In particular, we see that this truncation admits the $SO(8)$ invariant $AdS_4$ vacuum as well
as the unstable $SO(7)^+$ invariant one.

\subsubsection{2+2 equal charged scalars}
We next consider a different sub truncation of \eqref{popelag} 
that also keeps two charged scalar fields, two gauge fields and a single neutral scalar:
\begin{align}
\phi_{1}&=  \sigma,\qquad \phi_{2}=\phi_{3}  =0,\qquad\nn
\varphi_{1}&=\varphi_{2} =\gamma_{1},\qquad
\varphi_{3}=\varphi_{4}  =\gamma_{2}\nn
C^{1}&=C^{2}  =\frac{1}{\sqrt{2}}\,\left(A^{1}+\bar A^{0}\right),\qquad
C^{3}=-C^{4}  =\frac{1}{\sqrt{2}}\,\left(A^{1}-\bar A^{0}\right)\, .
\end{align}
The resulting equations of motion can be derived from an action with
Lagrangian
${\cal L}(g_{\mu\nu},\bar A^0_\mu, A^1_\mu,\gamma_1,\gamma_2,\sigma)$
given by
\begin{align}\label{2+2lag}
e^{-1}\mathcal{L}= & \frac{1}{2}R-\frac{1}{4}\left(\partial\sigma\right)^{2}-\frac{1}{2}\left(\partial\gamma_{1}\right)^{2}-\frac{1}{2}\left(\partial\gamma_{2}\right)^{2}-\frac{1}{2}\sinh^{2}\gamma_{1}\,\left(A^{1}+\bar A^{0}\right)^{2}\nn
&-\frac{1}{2}\sinh^{2}\gamma_{2}\,\left(A^{1}-\bar A^{0}\right)^{2}
  -\frac{1}{4}\cosh\left(\sigma\right)\,\left[\left(\bar F^{0}\right)_{\mu\nu}\left(\bar F^{0}\right)^{\mu\nu}+\left(F^{1}\right)_{\mu\nu}\left(F^{1}\right)^{\mu\nu}\right]\nn
&  -\frac{1}{2}\sinh\left(\sigma\right)\,\left[\left(F^{1}\right)_{\mu\nu}\left(\bar F^{0}\right)^{\mu\nu}\right]-V\, ,
 \end{align}
 with
\begin{align}
V= & -2\,\left(2\,\cosh\gamma_{1}\,\cosh\gamma_{2}+\cosh\sigma\right)\, .
\end{align}
For the maximally supersymmetric quantisation we have
\begin{align}
\Delta({\cal O}_{\gamma_1})=
\Delta({\cal O}_{\gamma_2})=
\Delta({\cal O}_{\sigma})=1\, ,
\end{align}
with $\gamma_i$ both having $U(1)_R$ charge 1 and $\sigma$ having $U(1)_R$ charge 0.

This truncation only has the $SO(8)$ invariant $AdS_4$ vacuum.
We also note that the truncation is invariant under the $\bbZ_2$ symmetry
\begin{align}\label{zed2}
\bar A^0\to -\bar A^0,\qquad \sigma\to -\sigma,\qquad \gamma_1\leftrightarrow\gamma_2\, .
\end{align}

\subsubsection{4 equal charged scalars}\label{4equal}
Finally we consider a different sub truncation\footnote{Note that there is an analogous truncation of $SO(6)$ gauged supergravity in $D=5$ that
was studied in \cite{Bhattacharyya:2010yg}.}
 that keeps a single charged scalar field and a single gauge field:
\begin{align}
\phi_{1}=\phi_{2}=\phi_{3} & =0\nn
\varphi_{1}=\varphi_{2}=\varphi_{3}=\varphi_{4} & =\varphi\nn
C^{1}=C^{2}=C^{3}=-C^{4} & =\frac{1}{\sqrt{2}}A^{1}\, .
\end{align}
The resulting equations of motion can be derived from an action with
Lagrangian
${\cal L}(g_{\mu\nu}, A^1_\mu,\varphi)$
given by
\begin{align}\label{4lag}
e^{-1}\mathcal{L}= & \frac{1}{2}R-\left(\partial\varphi\right)^{2}-\sinh^{2}\varphi\,\left(A^{1}\right)^{2}-\frac{1}{4}\,\left(F^{1}\right)_{\mu\nu}\left(F^{1}\right)^{\mu\nu}-V\nn
V= & -2\,\left(2+\cosh\left(2\varphi\right)\right)\, .
\end{align}
For the maximally supersymmetric quantisation we have
\begin{align}
\Delta({\cal O}_{\varphi})=1
\end{align}
and $\varphi$ has $U(1)_R$ charge 1.
This truncation only has the $SO(8)$ invariant $AdS_4$ vacuum.

Observe that this theory can be obtained from the 1+3 charged scalar truncation by setting $\phi=A^0=0$ and $\rho=\chi=\varphi$ in \eqref{1+3lag} 
and also from the 2+2 charged scalar truncation by setting $\sigma=\bar A^0=0$ and $\gamma_1=\gamma_2=\varphi$ in \eqref{2+2lag}. On the other hand
it cannot be obtained from the $SU(3)$ invariant truncation \eqref{eq:action}. 
Finally we observe that after setting $\varphi=0$ we obtain the Einstein-Maxwell
theory \eqref{emlag}.

\section{Perturbative instabilities of the $AdS-RN$ black brane}\label{secthree}
All of the truncations considered in the last section can be further truncated to the
Einstein-Maxwell theory given in \eqref{emlag}.
Thus all of the truncations admit the electrically charged AdS-RN black brane solution given by
\begin{align}\label{adsrn}
ds^{2}_{4}&=-f\,dt^{2}+\frac{dr^{2}}{f}+r^{2}\,\left(dx_{1}^{2}+dx_{2}^{2} \right)\nn
A^1&=\mu\,\left(1-\frac{r_{+}}{r}\right)dt\nn
f&=2r^{2}-\left(2r_{+}^{2}+\frac{\mu^{2}}{2} \right)\frac{r_{+}}{r}+\frac{\mu^{2}}{2}\frac{r_{+}^{2}}{r^{2}}\, .
\end{align}
The event horizon is located at $r=r_+$.
This describes the high temperature properties of the maximally supersymmetric $d=3$ CFT
when held at finite chemical potential $\mu$ with respect to the Reeb $U(1)_R$ gauge-field $A^1$.
In this section we first consider the linearised instabilities about this black brane solution within
the context of the truncated theories that we constructed in the last section.
It will be helpful to recall that at zero temperature, in the near horizon limit these solutions approach the $AdS_2\times \bbR^2$ solution 
\begin{align}\label{ads2sol}
ds^{2}_{4}&=-12r^{2}\,dt^{2}+\frac{dr^{2}}{12r^{2}}+\frac{\mu^2}{12}\,\left(dx_{1}^{2}+dx_{2}^{2}\right)\nn
F^1&=2\sqrt{3}\,dr\wedge dt\, .
\end{align}
We now set $\mu=1$ for convenience.

We next observe that all of the charged scalars in each of the truncated theories are minimally coupled and hence we can use the results\footnote{In the notation of \cite{Denef:2009tp} we have $M^{2}=1,\, L^{2}=1/2$, $g=1$ and hence $\gamma=1$.
Thus with our $\mu=1$ the critical temperatures $T_c$ that we give satisfy $T_c=\gamma T_c/\mu$. 
In  \cite{Denef:2009tp} some explicit values of $\gamma T_c/\mu$ for instabilities are also given but in comparing one should note that
they sometimes have $\gamma=1/2$.} 
of \cite{Denef:2009tp} to determine the critical temperatures at which they become unstable. First consider the $SU(3)$ invariant theory \eqref{eq:action}. As we have argued, in the maximal $d=3$ SCFT, 
$\zeta_2$, which has $U(1)_R$ charge 2, is dual to an operator with $\Delta=2$. From figure 1 of \cite{Denef:2009tp} this becomes
unstable at $T_{c}\approx 0.042$ (with $\mu=1$). 
The back reacted superfluid black branes were constructed in \cite{Gauntlett:2009dn,Gauntlett:2009bh}. 
However, we also see that $\zeta_1$, which has $U(1)_R$ charge 1, and is dual to an operator with $\Delta=1$ has a higher critical temperature at
$T_{c}\approx 0.174$. Thus we should focus on constructing black brane solutions with $\zeta_1\ne 0$. 
Similarly, the charged scalars in the truncations \eqref{1+3lag}, \eqref{2+2lag} and {\eqref{4lag}
all have $U(1)_R$ charge 1 and are dual to operators with $\Delta=1$ and have a critical temperature at
$T_{c}\approx 0.174$. 

We next consider the neutral scalar fields which are not minimally coupled and hence we cannot use the
results of \cite{Denef:2009tp}.  We again first consider the $SU(3)$ invariant theory \eqref{eq:action}. The linearised
fluctuations $\delta z$ couple with fluctuations of the gauge field strengths $\delta F^\alpha=d(\delta A^\alpha)$. 
We find that the real part of $\delta z$ and $\delta F^0$ decouple leading to
\begin{align}
(\Box+4-\frac{1}{2}F^{1}_{\mu\nu}F^{1\mu\nu})\delta\lambda+\frac{1}{2\sqrt{3}}\,F^{1}_{\mu\nu} \delta F^{0\mu\nu}&=0\nn
d\ast \delta F_{0}-2\sqrt{3}\, d\delta\lambda \wedge\ast F_{1}&=0\, .\label{GMeqs2}
\end{align}
Note that \eqref{GMeqs2} are precisely the same as those found by Gubser and Mitra in 
\cite{Gubser:2000ec}\cite{Gubser:2000mm}.
The linearised fluctuations of the imaginary part of $z$, $\delta\eta$, are a little more complicated. 
For simplicity, here we
restrict attention to fluctuations satisfying
\begin{align}
\delta F^{1}\wedge F^{1}=
d\delta\eta\wedge F^{1}=0\, .
\end{align}
The first condition means that we are not considering magnetic fluctuations of the gauge field while the second assumes that we are confining ourselves to translationally invariant fluctuations $\delta\eta$. We then obtain the equation of motion
\begin{equation}
(\Box+4+\frac{3}{2}F^1_{\mu\nu}F^{1\mu\nu})\delta\eta=0\, .\label{eq:eta_eom}
\end{equation}

We next consider the fluctuations of the neutral scalar field that appears in the truncations 
\eqref{1+3lag} and \eqref{2+2lag}. We again find that it couples to the additional gauge field 
and, up to relabelling, is again given by the Gubser-Mitra equations \eqref{GMeqs2}.

It is interesting to investigate these linearised equations in the near horizon limit at zero temperature i.e. 
in the $AdS_2\times \bbR^2$ background \eqref{ads2sol}.
We find that \eqref{eq:eta_eom} reads
\begin{equation}
(\Box_{AdS_{2}}-32)\delta\eta=0\, ,
\end{equation}
where $\Box_{AdS_{2}}$ is the scalar Laplacian on $AdS_2$ with radius squared $1/12$.
This satisfies the $AdS_2$ BF bound\footnote{Note that this is consistent, 
after using the identifications in footnote 4, 
with the dimension of the irrelevant operator used in equation $(5.14)$ of \cite{Gauntlett:2009bh} to deform the zero temperature limit of the AdS-RN black brane.} and suggests that there is no instability at finite temperature. In fact the stability of
this mode about the AdS-RN black brane solution at finite temperature was already observed in the analysis
of \cite{Gauntlett:2009bh}.  

We next consider the translationally invariant modes of the Gubser-Mitra system \eqref{GMeqs2} in the $AdS_2\times \bbR^2$ background. From equation \eqref{GMeqs2} we have that
\begin{align}
\delta F_{0}=2\sqrt{3} \delta\lambda\,F_{1}+k\, \mathrm{Vol}\left(AdS_{2}\right)\, ,
\end{align}
where $k$ is a constant. Since we are interested in the normalizable modes around the $AdS_2\times \bbR^2$ background we will set $k=0$. Substituting  back into \eqref{GMeqs2} we obtain
\begin{align}
(\Box_{AdS_{2}}-8)\delta\lambda&=0\, .
\end{align}
This satisfies $AdS_2$ bound and suggests these linearised modes are also stable in the 
AdS-RN black brane solution at finite temperature. However, this is not correct. Indeed, using a somewhat
indirect argument, Gubser and Mitra argued that the AdS-RN black brane should have an instability
\cite{Gubser:2000ec}\cite{Gubser:2000mm}. In fact we can demonstrate the existence of a critical 
temperature at which there  is a normalisable static mode that solves the system \eqref{GMeqs2}
using a shooting method.
Close to the black brane horizon we demand the smooth behaviour
\begin{align}
\delta\lambda=&\delta\lambda_{+}+\delta\lambda_{+}^{(1)}\,\left(r-r_{+}\right)+\dots\\
\delta A^{0}=&\delta \alpha_{+}^{(1)}\,\left(r-r_{+}\right)+\ldots\, ,
\end{align}
while at infinity we demand that there is zero deformation for the scalar field $\delta\lambda$ and no chemical potential for the one form $A^{0}$:
\begin{align}
\delta\lambda&=\frac{\delta\lambda^{(1)}}{r}+\frac{\delta\lambda^{(3)}}{r^{3}}\ldots\\
\delta A^{0}&=\frac{\delta q^{0}}{r}+\ldots\, ,
\end{align}
where we are quantising so that $\lambda$ is dual to an operator with $\Delta=1$.
We have checked that the critical temperature at which the zero modes exists is at $T_c\approx 0.159$ and we note that
$T_c=\frac{1}{2\pi}$ up to a discrepancy of the order of $10^{-6}$.

To summarise this section, we find that for all four truncations 
\eqref{su3lag}, \eqref{1+3lag}, \eqref{2+2lag} and {\eqref{4lag} 
the AdS-RN black brane solution \eqref{adsrn} has superfluid instabilities that set in at  $T_c \approx 0.174$ (with $\mu=1$)
and furthermore this is the highest critical temperature. 
The $SU(3)$ invariant truncation \eqref{su3lag} has an additional superfluid instability at $T_{c}\approx 0.042$ and is the one
associated with the superfluid black branes constructed in  \cite{Gauntlett:2009dn,Gauntlett:2009bh}.
The truncations with neutral scalars,
\eqref{su3lag}, \eqref{1+3lag} and \eqref{2+2lag}, also have Gubser-Mitra instabilities that set in 
at $T_{c}\approx 0.159$. In the next section we will construct the fully back reacted black brane solutions.
We will find a rich set of solutions and see that the different branches of black branes can become intricately
connected in the space of solutions.

\section{Back reacted black brane solutions}\label{sec4}
In this section we will construct various black brane solutions
for the four different truncations of
$SO(8)$ gauged supergravity that we obtained in section \ref{sectwo}. We are most interested
in constructing the black branes associated with the highest critical temperature, namely
the superfluid black branes associated with the condensation of the charged scalars dual to operators with $\Delta=1$
and $U(1)_R$ charge 1.
The simplest truncation is therefore {\eqref{4lag}, which contains just a metric, a single gauge field and a single charged
scalar and this will be our starting point in section \ref{41}. As we will see, the resulting superfluid black branes are rather
pathological in a somewhat surprising way. However, interestingly, in the other truncations that we then go on to study, 
we will find that there are different kinds of superfluid black brane solutions associated with the $\Delta=1$ instability that emanate from the AdS-RN black brane solution, 
as well as a rich variety of additional black branes solutions emanating from these new branches. We will also
see that the nature of the back reacted black branes associated with the Gubser-Mitra instability also depends on the truncation
and furthermore that they too can sprout additional superfluid black branes.
At the end of this section, we summarise our conclusions concerning the thermodynamics
within the truncations that we study.

In all cases we will consider the usual radial and purely electric ansatz with flat $\bbR^2$ spatial sections. 
For the metric we take
\begin{align}
ds_{4}^{2}&=-2r^{2}\,e^{-\beta\left(r\right)}g\left(r\right)\,dt^{2}+\frac{dr^{2}}{2r^{2}g\left(r\right)}+r^{2}\left(dx_{1}^{2}+dx_{2}^{2} \right)\, .\label{eq:rad_ansatz}
\end{align}
All scalar fields will just depend on $r$ and the charged scalars will be taken to be real. 
The $U(1)_R$ gauge field will take the form
$A^1=a_1(r) dt$ with an asymptotic expansion $a_1(r)=e^{-\beta_\infty}(\mu_1+q_1/r+\dots)$
where $\beta_\infty=\lim_{r\to \infty}\beta(r)$. 
The equations of motion are ODE's and there are two scaling symmetries which we will use to
set $\beta_\infty=0$ and $\mu_1=1$. For truncations where there is an additional gauge field we
will take a similar ansatz, but, in this section,  
will set the corresponding chemical potential to zero (this will be relaxed in section \ref{sec5}).
We will look for black brane solutions in which the other fields have an asymptotic falloff corresponding to the dual operator
acquiring a vev but no deformation. All of the black branes will be constructed using a shooting technique and
we refer to appendix C of \cite{Gauntlett:2009bh} for some details.

\subsection{4 equal charged scalars truncation} \label{41}

We begin by constructing superfluid black branes in the simplest truncation described 
by the Lagrangian ${\cal L}={\cal L}(g_{\mu\nu}, A^1_\mu,\varphi)$ given in \eqref{4lag}.
For the metric $g_{\mu\nu}$ we take the ansatz \eqref{eq:rad_ansatz}. For the single gauge field we take $A^1=a_1(r) dt$, while for the single
(real) charged scalar we take $\varphi=\varphi(r)$. The asymptotic expansion at infinity is given by 
\begin{align}\label{eq:SO8_exp}
g&=1+\frac{\varphi_{(1)}^{2}}{r^{2}}+\frac{m}{r^{3}}+\ldots\nn
\beta&=\frac{\varphi_{(1)}^{2}}{r^{2}}+\ldots
\nn
a_{1}&=\left(\mu_{1}+\frac{q_{1}}{r}\right)+\ldots\nn
\varphi&=\frac{\varphi_{(1)}}{r}+\frac{\varphi_{(2)}}{r^{2}}+\ldots\, .
\end{align}
As noted we can use a scaling symmetry to set $\mu_1=1$.
In order to remain within the quantisation appropriate for the
maximal supersymmetric $d=3$ CFT we impose boundary conditions so that $\Delta({\cal O}_\varphi)=1$. 
Thus the $1/r$ and $1/r^2$ falloffs of the scalar field $\varphi$
correspond in the dual CFT to the vev of the dual operator and a deformation of the Lagrangian by the dual operator, respectively.
The superfluid black branes that we construct will therefore have $\varphi_{(1)}\ne 0$ and $\varphi_{(2)}=0$.

We now consider the boundary conditions at the black brane horizon, which we assume is located at $r=r_+$. The expansion is taken to be
\begin{align}
g&= g^{(1)}_{+}\,\left(r-r_{+}\right)+\cdots\nn
\beta &= \beta_{+}+\beta^{(1)}_{+}\,\left(r-r_{+}\right)+\cdots\nn
a_{1} &= a_{1+}^{(1)}\,\left(r-r_{+}\right)+\cdots\nn
\varphi &= \varphi_{+}+\varphi^{(1)}_{+}\,\left(r-r_{+} \right)+\cdots
\end{align}
and one can show the expansion is specified by $\beta_{+},a_{1+}^{(1)}, \varphi_{+}$.
The Hawking temperature of the black brane is given by $T=r^{2}_{+}g^{\prime}\left(r_{+}\right)e^{-\beta\left(r_{+}\right)/2}/\left(2\pi\right)$. The entropy density is given by $s=2\pi r_+^2$ and the free energy density $w=w(T,\mu_1,\varphi_2)$ 
is given by $w=m$ and we also have the Smarr formula $3m=-sT+q_{1}\mu_{1}+8\varphi_{(1)}\varphi_{(2)}$, which for the configurations of interest reads $3m=-sT+q_{1}\mu_{1}$.
A representative calculation leading to these results is included in appendix A for the more complicated $SU(3)$ invariant truncation.

We now construct superfluid black branes.
We need to solve first order ODE's for $g,\beta$ and second order ODE's
for $\varphi,a_1$ and thus we need to specify six parameters to specify a unique solution.
With the scaling symmetries fixed, we 
have seven parameters $r_+,\beta_+,a_{1+}^{(1)},\varphi_+,m,q_1,\varphi_1$ to specify which implies that we should
expect a one parameter family of solutions (assuming they exist). 
We take this parameter to be the temperature of the black branes.

We look for these black branes using a shooting technique 
and do indeed find a new branch of superfluid black branes emanating from the AdS-RN black branes at $T_c\approx 0.174$ as expected.
What is unexpected, however, is that the superfluid black brane solutions appear to only exist for temperatures higher than
$T_c$. In figure \ref{fig:charged_VEVs_single} 
we have plotted the asymptotic value $\varphi_{(1)}$, which is proportional to the vev of the order parameter
$<{\cal O}_\varphi>$, and indeed our numerics indicate that the branch may well continue 
all the way out to infinite temperatures; in particular, we don't find
any evidence of the branch turning back to lower temperatures as characteristic of a first order transition. 
In figure \ref{fig:charged_VEVs_single} we have plotted the free energy of these new superfluid black branes and, reassuringly, 
we find that it is always greater than that of the AdS-RN black brane
and so they are never thermodynamically preferred. 
These superfluid black branes thus provide the first string/M-theory 
embeddings of ``exotic hairy black holes" that were constructed in \cite{Buchel:2009ge},
building on \cite{Gubser:2005ih}. We also point out that
we have found no evidence for any other black branes emanating from the superfluid branch. To investigate this issue we
use the method described in appendix C of \cite{Gauntlett:2009bh}.

\begin{figure}
\centering
\subfloat[A plot of $\varphi_{(1)}\sim <{\cal O}_\varphi>$ versus $T$]{\includegraphics[width=7cm]{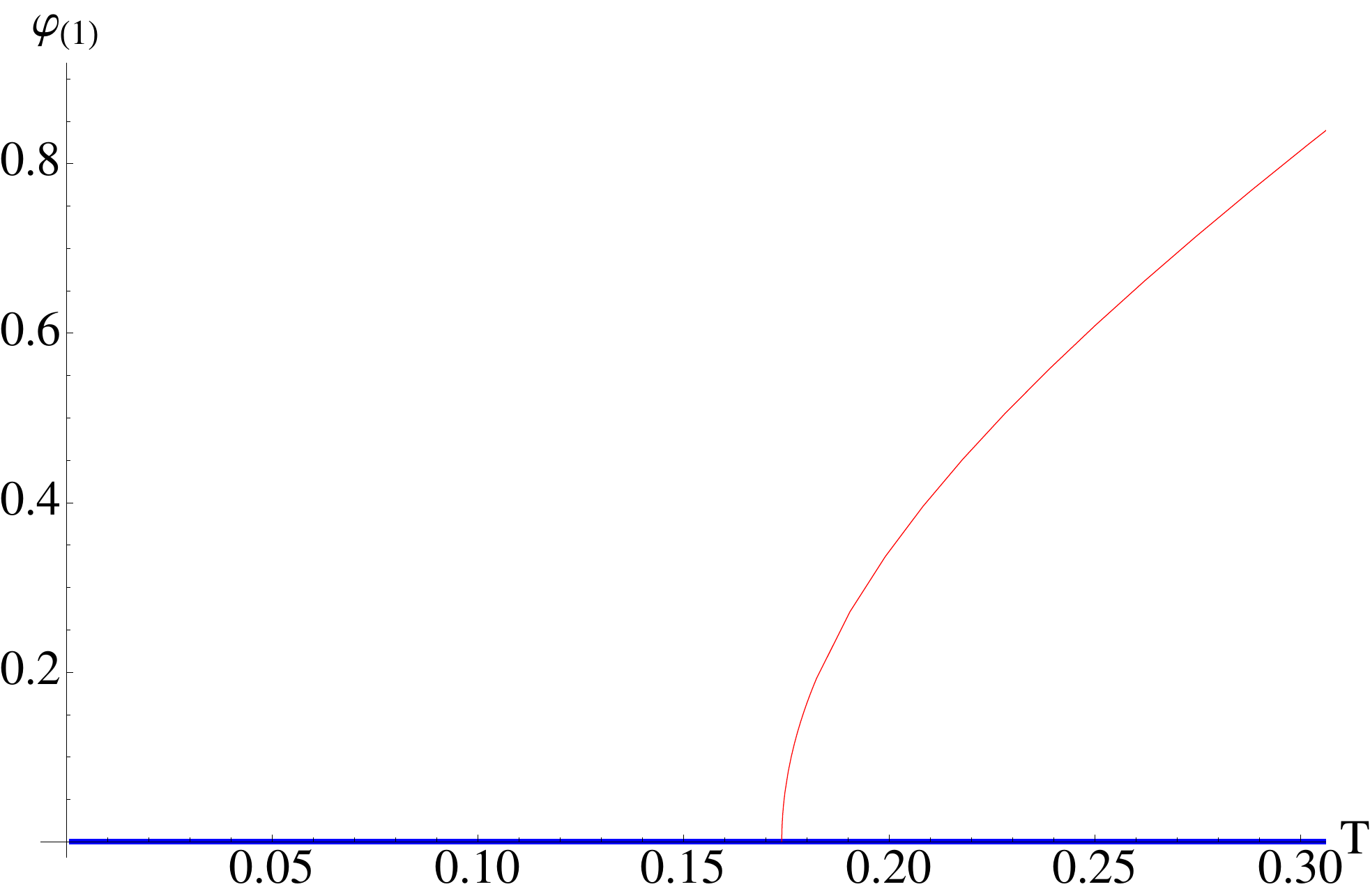}}
\subfloat[A plot of the free energy density versus $T$]{\includegraphics[width=7cm]{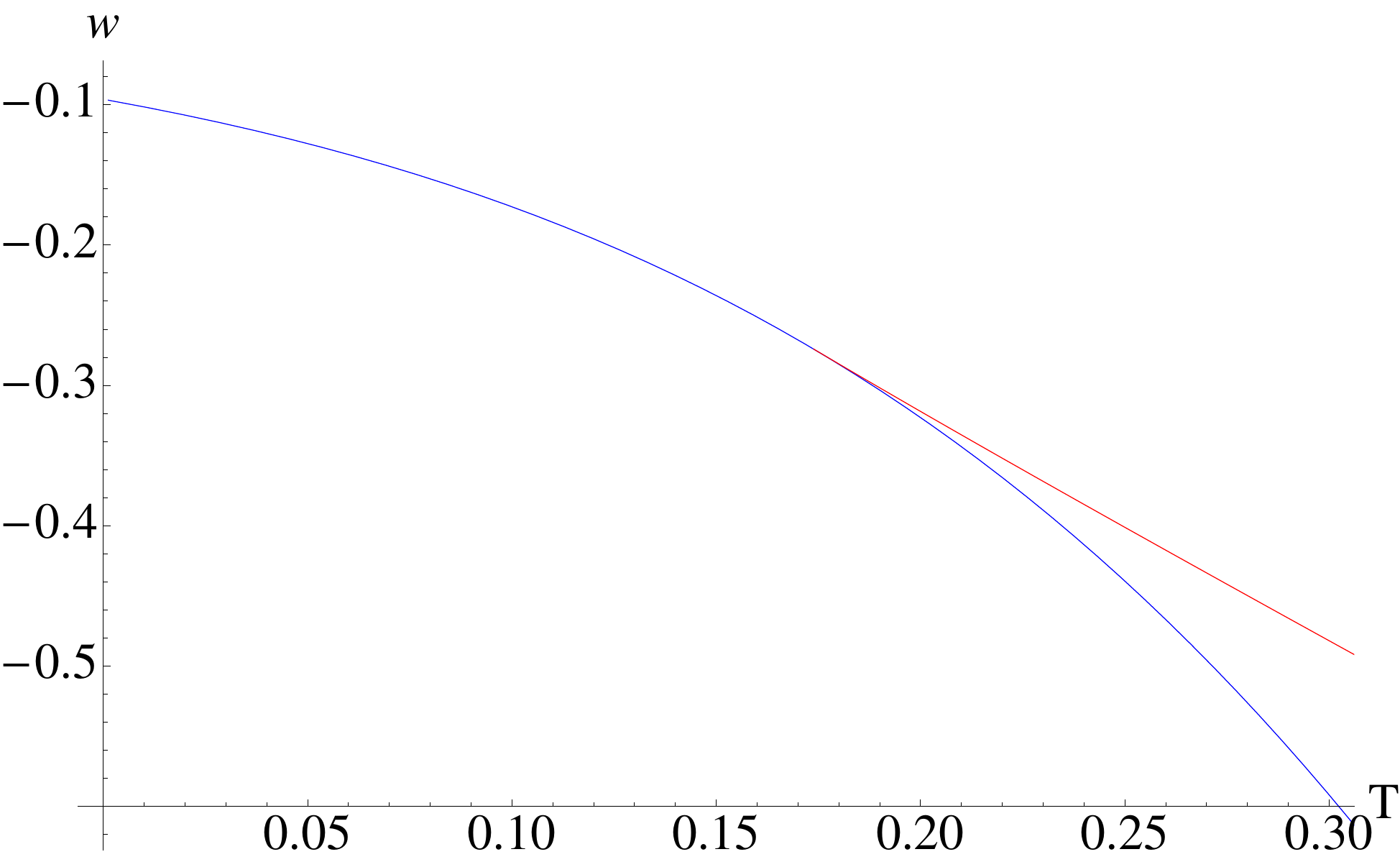}}
\caption{
Black brane solutions in the 4 equal charged scalars truncation \eqref{4lag} with $\mu_1=1$.
The plots show $\varphi_{(1)}\sim <{\cal O}_\varphi>$ and the free energy $w$
versus temperature $T$ for the superfluid black brane solutions (red line) emanating from the
AdS-RN black brane solutions (blue line) at $T_c\approx 0.174$.}
\label{fig:charged_VEVs_single}
\end{figure}


At this juncture one might be tempted to think that these rather pathological black branes are generic for the 
$\Delta=1$ charged scalars in $SO(8)$ gauged supergravity. However, we will see very different types of superfluid
black branes associated with $\Delta=1$ charged scalars
within other truncations. Thus as far as $SO(8)$ gauged supergravity is concerned the new superfluid 
black branes of this truncation are not physically relevant and appear to be a curiosity. 
Nevertheless, we think it would be interesting to further study the dynamical behaviour of the AdS-RN black branes within this present
truncation at temperatures below $T_c$ as non-translationally invariant black branes 
are likely to play an important role.

\subsection{2+2 equal charged scalars truncation}

We next construct superfluid and other black brane solutions in the truncation described by the Lagrangian 
${\cal L}={\cal L}(g_{\mu\nu}, A^1_\mu,\bar A^0_\mu,\gamma_1,\gamma_2,\sigma)$ given in \eqref{2+2lag}.
For the metric ansatz we again take \eqref{eq:rad_ansatz} and for the gauge fields we take  $A^1=a_1(r) dt$ and
$\bar A^0=\bar a_0(r) dt$. The charged scalars $\gamma_1$ and $\gamma_2$ and the neutral scalar $\sigma$ are functions of $r$.
The asymptotic expansion at infinity is given by
\begin{align}\label{eq:SO8_exp2}
g&=1+\frac{1}{2r^{2}}\,\left(\gamma_{1(1)}^{2}+\gamma_{2(1)}^{2}+\frac{1}{2}\sigma_{(1)}^{2} \right)+\frac{m}{r^{3}}+\ldots\nn
\beta&=\frac{1}{2r^{2}}\,\left(\gamma_{1(1)}^{2}+\gamma_{2(1)}^{2}+\frac{1}{2}\sigma_{(1)}^{2} \right)+\ldots
\nn
a_{1}&=\left(\mu_{1}+\frac{q_{1}}{r}\right)+\ldots,\qquad
\bar a_{0}=\left(\bar\mu_0+\frac{\bar q_{0}}{r}\right)+\ldots\nn
\gamma_i&=\frac{\gamma_{i(1)}}{r}+\frac{\gamma_{i(2)}}{r^{2}}+\ldots,\qquad \sigma=\frac{\sigma_{(1)}}{r}+\frac{\sigma_{(2)}}{r^{2}}+\ldots\, .
\end{align}
We will again often scale so that $\mu_1=1$. In the black brane solutions that we construct we will 
set the other chemical potential $\bar\mu_0=0$ but will allow for 
a charge $\bar q_0$ which corresponds to the dual operator ${\cal O}_{\bar A^0}$ acquiring a vev. 
In the quantisation appropriate for the
maximal supersymmetric $d=3$ CFT the scalars are all dual to operators with $\Delta=1$ 
and so the $1/r$ and $1/r^2$ falloffs correspond in the dual CFT to the vev of the dual operator and a deformation of the Lagrangian by the dual operator, respectively. In all of the black branes we construct we will thus
set $\gamma_{i(2)}=\sigma_{(2)}=0$. 
Solutions with $\gamma_{1(1)}\ne 0$ or $\gamma_{2(1)}\ne 0$, corresponding
to $<{\cal O}_{\gamma_1}>\ne0$ or $<{\cal O}_{\gamma_2}>\ne0$, will spontaneously
break the $U(1)\times U(1)$ symmetry to a $U(1)$ and if both are non-zero there will be no residual symmetry.

We now consider the boundary conditions at the black brane horizon, which we assume is located at $r=r_+$. The expansion is taken to be
\begin{align}
g&= g^{(1)}_{+}\,\left(r-r_{+}\right)+\cdots,\qquad\qquad
\beta = \beta_{+}+\beta^{(1)}_{+}\,\left(r-r_{+}\right)+\cdots\nn
a_{1} &= a_{1+}^{(1)}\,\left(r-r_{+}\right)+\cdots, \qquad\qquad
\bar{a}_{0} = \bar a_{0+}^{(1)}\,\left(r-r_{+}\right)+\cdots\nn
\gamma_i &= \gamma_{i+}+\gamma^{(1)}_{i+}\,\left(r-r_{+} \right)+\cdots,\qquad  \sigma = \sigma_{+}+\sigma^{(1)}_{+}\,\left(r-r_{+} \right)+\cdots
\end{align}
and one can show the expansion is specified by $\beta_{+},a_{1+}^{(1)}, \bar a_{1+}^{(1)},\gamma_{i+},\sigma_+$.
The Hawking temperature of the black brane is given by $T=r^{2}_{+}g^{\prime}\left(r_{+}\right)e^{-\beta\left(r_{+}\right)/2}/\left(2\pi\right)$. The entropy density by $s=2\pi r_+^2$ and the free energy density 
$w=w(T,\mu_1,\gamma_{i(2)},\sigma_{(2)})$ is given by $w=m$. We also have the 
Smarr formula $3m=-sT+q_{1}\mu_{1}+4\gamma_{1(1)}\gamma_{1(2)}+4\gamma_{2(1)}\gamma_{2(2)}+2\sigma_{(1)}\sigma_{(2)}$, which, for the configurations of interest, reads $3m=-sT+q_{1}\mu_{1}$.

We now construct black branes emerging from the AdS-RN black brane solution
using the shooting technique that we described in the last subsection. In particular, a simple parameter count 
again suggests that the type of black branes that we are looking for can again be parametrised by their temperature.
To explain our results, we first recall that upon setting 
$\sigma=\bar A^0=0$ and $\gamma_1=\gamma_2=\varphi$ we recover the 4 equal charged scalar truncation
and so the superfluid black branes constructed in the last subsection (see figure \ref{fig:charged_VEVs_N12}) 
are also solutions in the present truncation. Given that these superfluid black branes are now living in a larger theory
there could in principle be additional branches of black brane solutions that emanate from them. However, we have checked
and find no evidence for them.

We now look for additional black brane solutions that emanate from the AdS-RN black brane solution. 
In the following it will be helpful to 
recall that this model has the $\bbZ_2$ symmetry \eqref{zed2}. In addition to the superfluid black branes appearing at $T_c\approx 0.174$ with 
$\gamma_1=\gamma_2$ and $\sigma=0$, there are also two other branches of superfluid black branes that appear at the same temperature.
The first has $\gamma_1\ne0$ and $\gamma_2=0$ and also $\sigma\ne 0, \bar A^0\ne 0$ while the second one is related by
the $\bbZ_2$ symmetry and has, in particular,
$\gamma_2\ne0$ and $\gamma_1=0$. 
Interestingly, unlike those of figure \ref{fig:charged_VEVs_N12}, these two branches of superfluid black branes exist
for temperatures less than $T_c$ and moreover are thermodynamically preferred over the AdS-RN black branes. In figures
\ref{fig:charged_VEVs_N12}(a) and \ref{fig:charged_VEVs_N12}(b)
we have plotted the
value of $\gamma_{1(1)}$, that gives the vev $<{\cal O}_{\gamma_1}>$, and 
$\sigma_{1(1)}$, that gives the vev $<{\cal O}_{\sigma_1}>$, against temperature for the superfluid black brane branch with 
$\gamma_2=0$ as solid red lines. 

\begin{figure}
\centering
\subfloat[A plot of $\gamma_{1(1)}\sim <{\cal O}_{\gamma_1}>$ versus temperature $T$]{\includegraphics[width=10cm]{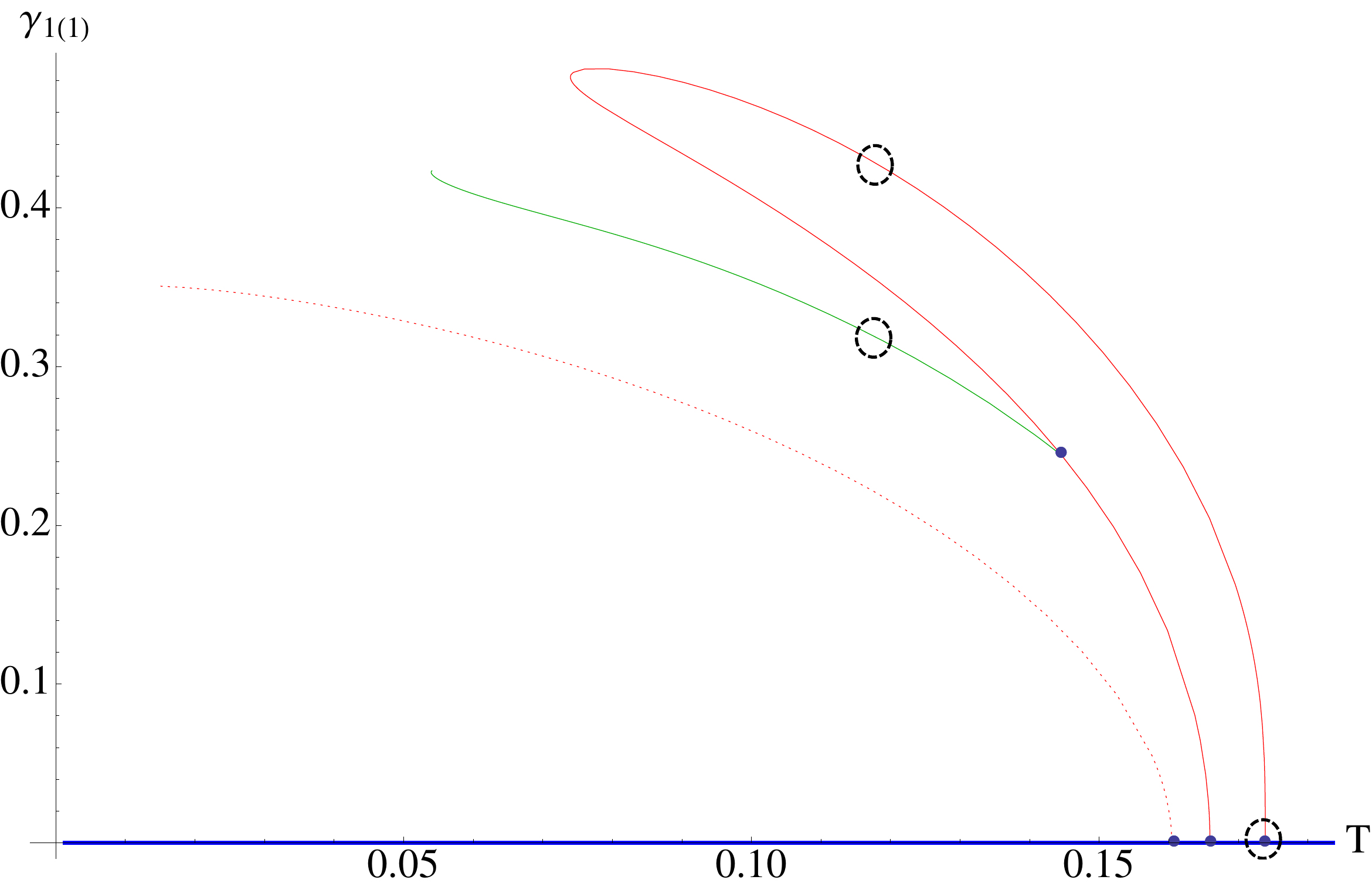}}
\\
\subfloat[A plot of  $\sigma_{(1)}\sim <{\cal O}_{\sigma}>$ versus temperature $T$]{\includegraphics[width=10cm]{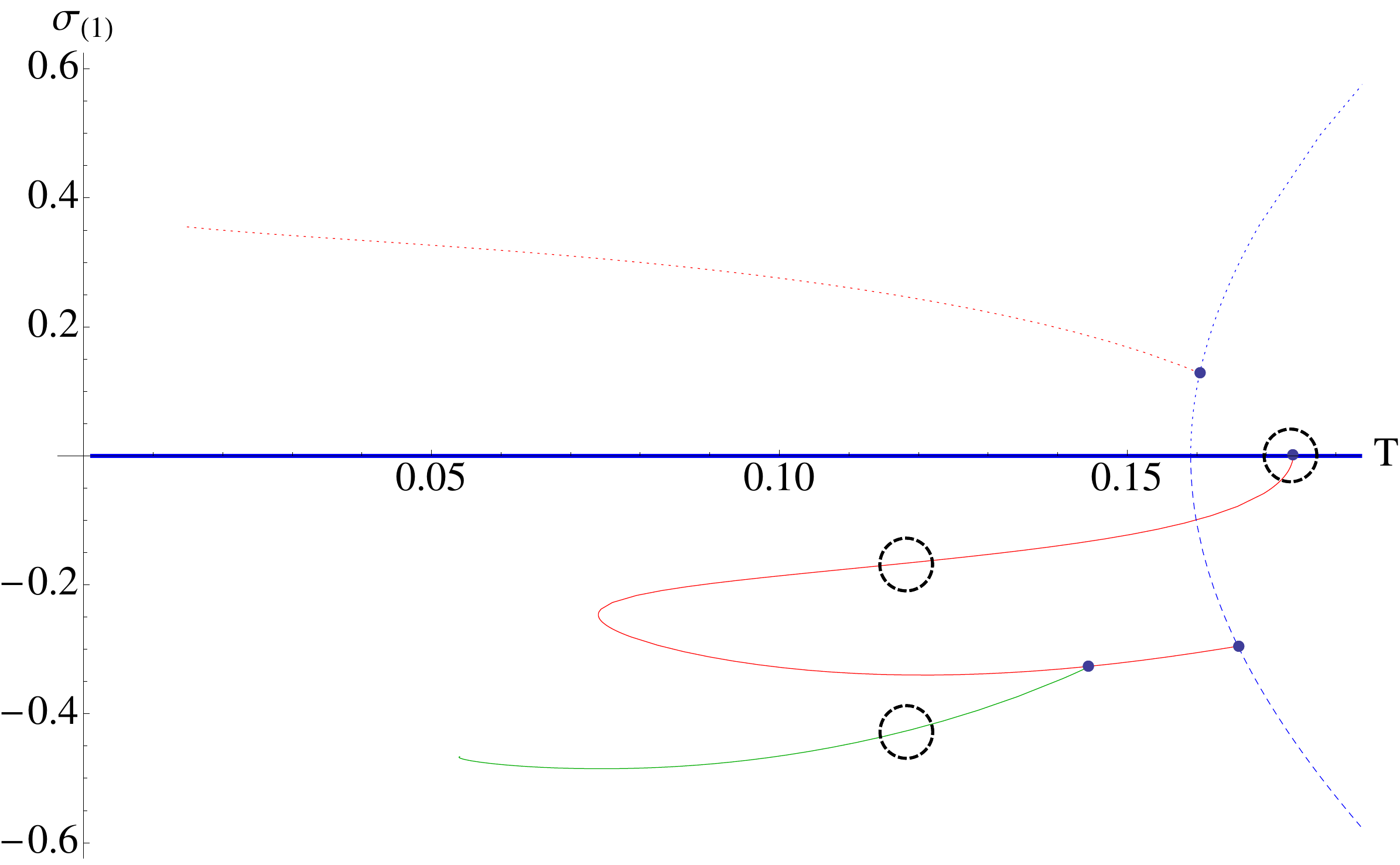}}
\caption{Various black brane solutions in the 2+2 equal charged scalars truncation \eqref{2+2lag} with $\mu_1=1$, $\bar\mu_0=0$:
additional solutions are found using the $\bbZ_2$ symmetry \eqref{zed2}.
The red branches are superfluid black branes with $\gamma_1\ne 0$ and $\gamma_2=0$. The solid red branch is the superfluid black brane associated with an
instability of the AdS-RN black brane. The green line is a branch of superfluid black branes that in addition
has $\gamma_{2}\ne 0$.
The blue dotted and dashed lines are two branches of black branes with $\gamma_1=\gamma_2=0$
associated with the Gubser-Mitra instability of the AdS-RN black brane.
The solid blue line in the bottom figure is the AdS-RN black brane, but in the top figure it also represents the back reacted Gubser-Mitra 
branes.
The solid dots on lines that meet indicate the points at which a new branch of black branes is appearing.
The dashed circles indicate the
thermodynamically preferred phase transitions within this truncation: as one lowers $T$
the system moves from the AdS-RN branch to the solid red superfluid
branch, then, discontinuously, to the green branch.}\label{fig:charged_VEVs_N12}
\end{figure}

Interestingly we find that this solid red line branch exists up to some minimal temperature before looping back to higher temperatures, ending on 
a new branch, which we will return to in a moment. Along the superfluid solid red line branch there is also a new branch of different
superfluid black branes that emanate at $T_c\approx 0.145$ which, unlike the solid red line branch, also has $\gamma_{2(1)}\ne 0$ and
hence $<{\cal O}_{\gamma_2}>\ne 0$. These superfluid black branes are marked on 
figures \ref{fig:charged_VEVs_N12}(a) and \ref{fig:charged_VEVs_N12}(b)
as a solid green line. Another interesting feature is that the solid green line branch seems to exist only down to $T\approx 0.054$ 
where the solution appears to become nakedly singular. Note that all of the black branes with $\sigma\ne 0$, for example all of 
the black branes with $\zeta_1\ne0$, lead to $\bar A^0\ne 0$ and more specifically $\bar q_0\ne 0$.

We also find that there are additional branches of black brane solutions that emanate from the AdS-RN black brane solutions. 
Specifically, associated with the Gubser-Mitra critical instability at temperature $T_c\approx 0.159 $ that we discussed in the last section,
we find two branches of black branes appearing, again related by the $\bbZ_2$ symmetry, with
$\gamma_1=\gamma_2=0$ but $\sigma\ne 0$, $\bar A^0\ne 0$. These black branes are together marked 
on 
figures \ref{fig:charged_VEVs_N12}(a) and \ref{fig:charged_VEVs_N12}(b)
as a dotted and dashed blue lines. We see that these black branes again appear to have the strange feature of 
only existing at temperatures great than $T_c$. They have higher free energy than the AdS-RN black branes and so are not thermodynamically relevant.

It is interesting that there are other branches of black brane solutions which emanate from the Gubser-Mitra branches. 
One is a superfluid black brane 
with $\gamma_2=0$ but 
$<{\cal O}_{\gamma_1}>\ne 0$, $<{\cal O}_{\sigma}>\ne 0$ and $<{\cal O}_{\bar A^0}>\ne 0$ (i.e. $\bar q_0\ne 0$) and is
marked with a dotted red line on figures \ref{fig:charged_VEVs_N12}(a) and \ref{fig:charged_VEVs_N12}(b), and there is another related by the $\bbZ_2$ symmetry. Furthermore, we find that the solid red line branch of superfluid black brane solutions ends on the Gubser-Mitra branch
as shown in figures \ref{fig:charged_VEVs_N12}(a) and \ref{fig:charged_VEVs_N12}(b).

By analysing the free energy of all of the new black brane solutions we find, within this truncation,
the following thermodynamic picture. As one lowers the temperature there is a second order phase transition
at $T\approx0.174$ from the AdS-RN phase to the superfluid phase marked by a solid red line in figures
\ref{fig:charged_VEVs_N12}(a) and \ref{fig:charged_VEVs_N12}(b) which breaks the $U(1)\times U(1)$ symmetry to $U(1)$. 
One stays in this phase down to
$T\approx 0.118$ at which point there is a first order phase transition to a new superfluid branch
marked by the solid green line 
in figures \ref{fig:charged_VEVs_N12}(a) and \ref{fig:charged_VEVs_N12}(b) which then breaks the residual $U(1)$ symmetry.
We can follow this phase down until $T\approx 0.054$ at which point the black branes become nakedly singular.

\subsection{1+3 equal charged scalars truncation}
We next consider the truncation described by the Lagrangian 
${\cal L}={\cal L}(g_{\mu\nu}, A^1_\mu, A^0_\mu,\rho,\chi,\phi)$ given in \eqref{1+3lag}.
For the metric ansatz we again take \eqref{eq:rad_ansatz} and for the gauge fields we take  $A^1=a_1(r) dt$ and
$A^0=a_0(r) dt$. The charged scalars $\rho$ and $\chi$ and the neutral scalar $\phi$ are functions of $r$.
The asymptotic expansion at infinity is given by
\begin{align}\label{eq:SO8_exp3}
g&=1+\frac{1}{4r^{2}}\left(\rho_{(1)}^{2}+3\,\chi_{(1)}^{2}+3\,\phi_{(1)}^{2} \right)+\frac{m}{r^{3}}+\ldots\nn
\beta&=\frac{1}{4r^{2}}\left(\rho_{(1)}^{2}+3\,\chi_{(1)}^{2}+3\,\phi_{(1)}^{2} \right)+\ldots
\nn
a_{1}&=\left(\mu_{1}+\frac{q_{1}}{r}\right)+\ldots,\qquad
a_{0}=\left(\mu_0+\frac{ q_{0}}{r}\right)+\ldots\nn
\rho&=\frac{\rho_{(1)}}{r}+\frac{\rho_{(2)}}{r^{2}}+\ldots,\qquad
\chi=\frac{\chi_{(1)}}{r}+\frac{\chi_{(2)}}{r^{2}}+\ldots,\qquad \phi=\frac{\phi_{(1)}}{r}+\frac{\phi_{(2)}}{r^{2}}+\ldots\, .
\end{align}
We are interested in black branes with $\mu_1\ne 0, \mu_0=0$ and also $\rho_{(2)}=\chi_{(2)}=\phi_{(2)}=0$
corresponding to the dual operators with $\Delta=1$ in the maximally supersymmetric $d=3$ CFT just
acquiring vev's.

When $\chi=0$ this truncation gives the sub-truncation \eqref{su3lag} of the $SU(3)$ invariant sector with $\zeta_2=0$. 
Since we will discuss in detail the black brane solutions of the latter theory in the next section we will not duplicate a discussion of
the relevant solutions here. One potentially important point is that when the  solutions are considered in the context of
the 1+3 equal charged scalars truncation 
 there could be additional solutions branching off 
from these. 
However, we have found that the only new branches of solutions that exist 
in the 1+3 truncation that are not in the $SU(3)$ truncation 
are both superfluid branches. One is the superfluid branch that we already discussed in the 4 equal charged scalar
truncation in section 4.1. The other has
$\rho_{(1)}=0$ but $\chi_{(1)}\ne 0$, $\phi_{(1)}\ne 0$, $q_0\ne 0$. The behaviour of the 
$\chi_{(1)}\ne 0$ versus temperature is very much like that in figure \ref{fig:charged_VEVs_single} and has greater free energy
than that of the AdS-RN black branes.

\subsection{The $SU(3)$ truncation}
We next consider superfluid and other black brane solutions in the truncation described by the Lagrangian 
${\cal L}={\cal L}(g_{\mu\nu}, A^1_\mu, A^0,\zeta_1,\zeta_2,\lambda)$ given in \eqref{su3lag}. 
For the metric ansatz we again take \eqref{eq:rad_ansatz} and for the gauge fields we take  $A^1=a_1(r) dt$ and
$A^0=a_0(r) dt$. The charged scalars $\zeta_i$, which we take to be real,  
and the neutral scalar $\lambda$ are functions of $r$.
The asymptotic expansion at infinity is given by 
\begin{align}\label{eq:SO8_exp4}
g&=1+\frac{1}{r^{2}}\,\left(\zeta_{1(1)}^{2}+\zeta_{2(1)}^{2}+3\lambda_{(1)}^{2} \right)+\frac{m}{r^{3}}+\cdots\nn
\beta&=\frac{1}{r^{2}}\,\left(\zeta_{1(1)}^{2}+\zeta_{2(1)}^{2}+3\lambda_{(1)}^{2} \right)+\cdots\ldots
\nn
a_{1}&=\left(\mu_{1}+\frac{q_{1}}{r}\right)+\ldots,\qquad
a_{0}=\left(\mu_0+\frac{q_{0}}{r}\right)+\ldots\nn
\zeta_i&=\frac{\zeta_{i(1)}}{r}+\frac{\zeta_{i(2)}}{r^{2}}+\ldots,\qquad \lambda=\frac{\lambda_{(1)}}{r}+\frac{\lambda_{(2)}}{r^{2}}+\ldots\, .
\end{align}
We will again often scale so that $\mu_1=1$. In this section we will be interested in black branes with the other chemical potential
set to zero, $\mu_0=0$, but we will allow for a charge $\bar q_0$ which corresponds to the dual operator ${\cal O}_{\bar A^0}$ acquiring a vev. 
In the quantisation appropriate for the maximal supersymmetric $d=3$ CFT we have $\Delta({\cal O}_{\zeta_1})
= \Delta({\cal O}_{\lambda})=1$ while $\Delta({\cal O}_{\zeta_2})=2$. Thus we will be interested in black brane solutions with
$\zeta_{1(2)}=\lambda_{(2)}=0$ and $\zeta_{2(1)}=0$. Solutions with $\zeta_{1(1)}\ne 0$ or $\zeta_{2(2)}\ne0$, corresponding
to $<{\cal O}_{\zeta_1}>\ne0$ or $<{\cal O}_{\zeta_2}>\ne0$, respectively, will spontaneously
break the $U(1)\times U(1)$ symmetry to a $U(1)$ and if both are non-zero there will be no residual symmetry.


We now consider the boundary conditions at the black brane horizon, which we assume is located at $r=r_+$. The expansion is taken to be
\begin{align}
g&= g^{(1)}_{+}\,\left(r-r_{+}\right)+\cdots,\qquad\qquad
\beta = \beta_{+}+\beta^{(1)}_{+}\,\left(r-r_{+}\right)+\cdots\nn
a_{1} &= a_{1+}^{(1)}\,\left(r-r_{+}\right)+\cdots, \qquad\qquad
{a}_{0} = a_{0+}^{(1)}\,\left(r-r_{+}\right)+\cdots\nn
\zeta_i &= \zeta_{i+}+\zeta^{(1)}_{i+}\,\left(r-r_{+} \right)+\cdots,\qquad  \sigma = \sigma_{+}+\sigma^{(1)}_{+}\,\left(r-r_{+} \right)+\cdots,
\end{align}
and one can show the expansion is specified by $\beta_{+},a_{1+}^{(1)}, a_{0+}^{(1)},\zeta_{i+},\sigma_+$.
The Hawking temperature of the black brane is given by $T=r^{2}_{+}g^{\prime}\left(r_{+}\right)e^{-\beta\left(r_{+}\right)/2}/\left(2\pi\right)$. The entropy density by $s=2\pi r_+^2$ and the free energy density 
$w=w(T,\mu_1,\zeta_{1(2)},\zeta_{(1)},\lambda_{(2)})$ is given by $w=m-4 \zeta_{2(1)}\zeta_{2(2)}$ which gives $w=m$
for the solutions that we are studying. We also have the 
Smarr formula $3m=-sT+\mu_{0}q_{0}+\mu_{1}q_{1}+24\lambda_{(1)}\lambda_{(2)}+8\zeta_{1(1)}\zeta_{1(2)}+8\zeta_{2(1)}\zeta_{2(2)}$.

We now construct black branes emerging from the AdS-RN black brane solution
using the shooting technique that we described earlier. 
We find a rich array of solutions and for each of them we have plotted
the values of $\lambda_{(1)}$,  $\zeta_{1(1)}$ and $\zeta_{2(2)}$ against temperature  
in figure \ref{su3fig}. In figure \ref{su3fig}(a) the blue branch on the horizontal axis corresponds to the AdS-RN black brane.
At $T_c\approx 0.174$ the solid red line emerges from the AdS-RN branch and 
are the superfluid black branes associated with the $\Delta=1$ charged scalar $\zeta_1$. As one lowers
the temperature along the solid red line branch one finds a second branch of superfluid black 
branes that also have $\zeta_2\ne 0$ at $T_c\approx 0.0297$, labelled as a solid green line. 

The dotted and dashed blue branches emerging from the AdS-RN black branch at $T_c\approx 0.159$ 
are associated with the Gubser-Mitra instability. 
In contrast to the analogous black branes found in the 2+2 truncation (see figure \ref{fig:charged_VEVs_N12}), we see in
figure \ref{su3fig}(a) that these black branes
exist for temperatures lower than the branching temperature and we can also show that they are now thermodynamically preferred over the AdS-RN black branes. 
Once again we observe that 
the details of the truncation can significantly affect the nature of the back reacted black brane solutions.

Note that in figure \ref{su3fig}(b) and figure \ref{su3fig}(c) the blue branch along the horizontal axis represents both the AdS-RN black brane and
the Gubser-Mitra black branes. Figure \ref{su3fig} also show some other superfluid black branes emerging from the Gubser-Mitra branches which in turn branch again. 

In figure \ref{su3fig}(c) we also see superfluid black branes with $\zeta_1=\sigma=0$ emerging form the AdS-RN branch 
at $T_c\approx 0.042$, labelled as a solid black branch. These are precisely the black branes found in \cite{Gauntlett:2009dn} that at zero temperature are smooth domain wall solutions interpolating between the $SO(8)$ and
the $SU(4)^-$ $AdS_4$ vacua. 
Unlike the truncations of \cite{Gauntlett:2009dn}, \cite{Gauntlett:2009bh},
in the truncation
of this paper we see that the solid black line branches into two purple superfluid branches with $\zeta_1=0$ but $\sigma\ne 0$.

By analysing the free energy of all of the different branches we find that at
$T_c\approx 0.174$ the system moves from the AdS-RN branch, preserving $U(1)\times U(1)$ to 
the solid red branch which preserves a $U(1)$. Then at $T_{c}\approx 0.132$ there is a first order transition to the dotted Gubser-Mitra branch. As far as we can tell this branch appears to become nakedly singular at zero temperature.

\begin{figure}
\centering
\subfloat[A plot of $\lambda_{(1)}\sim <{\cal O}_{\lambda}>$ versus temperature $T$]{\includegraphics[width=10cm]{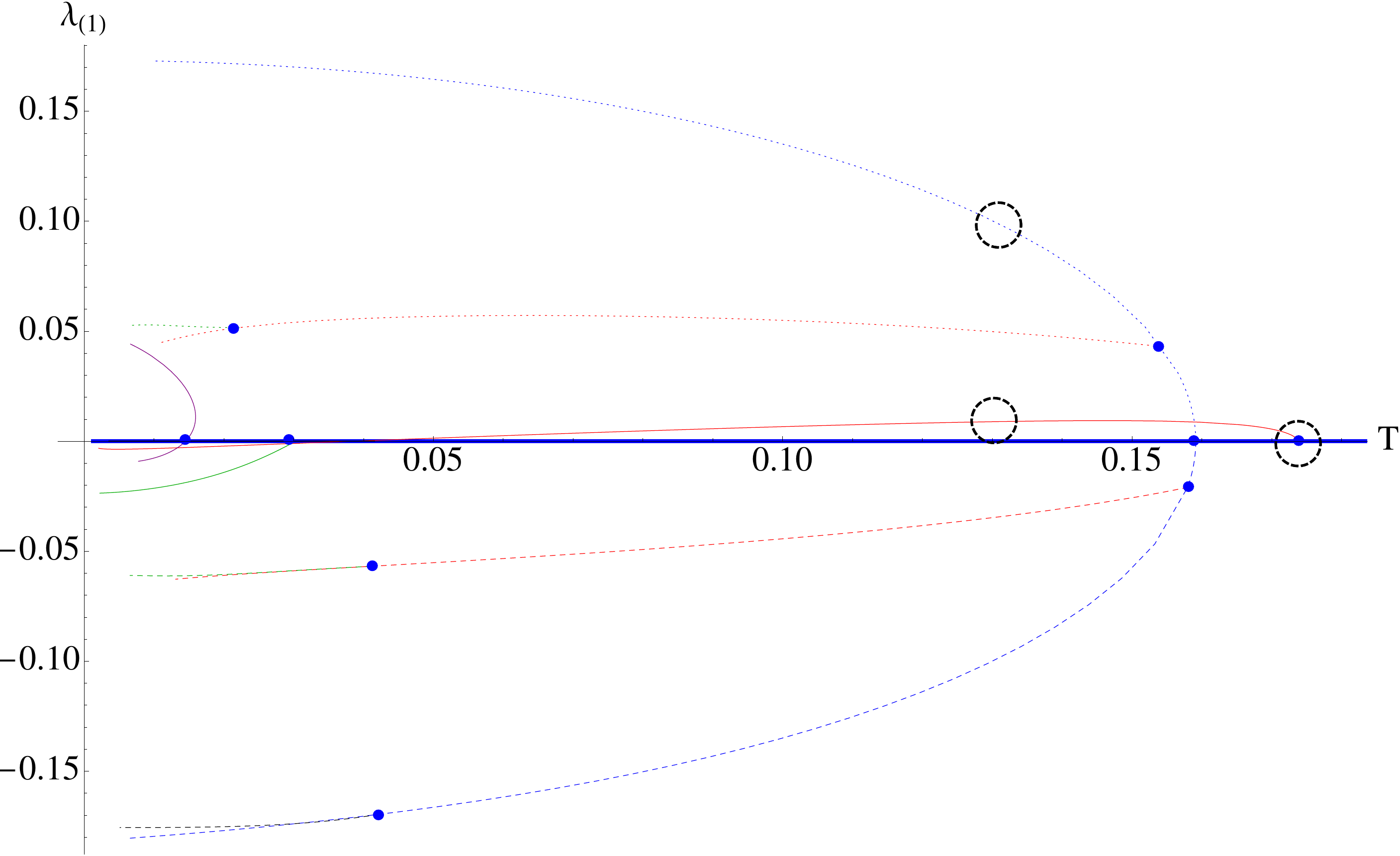}}\\
\subfloat[A plot of $\zeta_{1(1)}\sim <{\cal O}_{\zeta_1}>$ versus $T$]{\includegraphics[width=8cm]{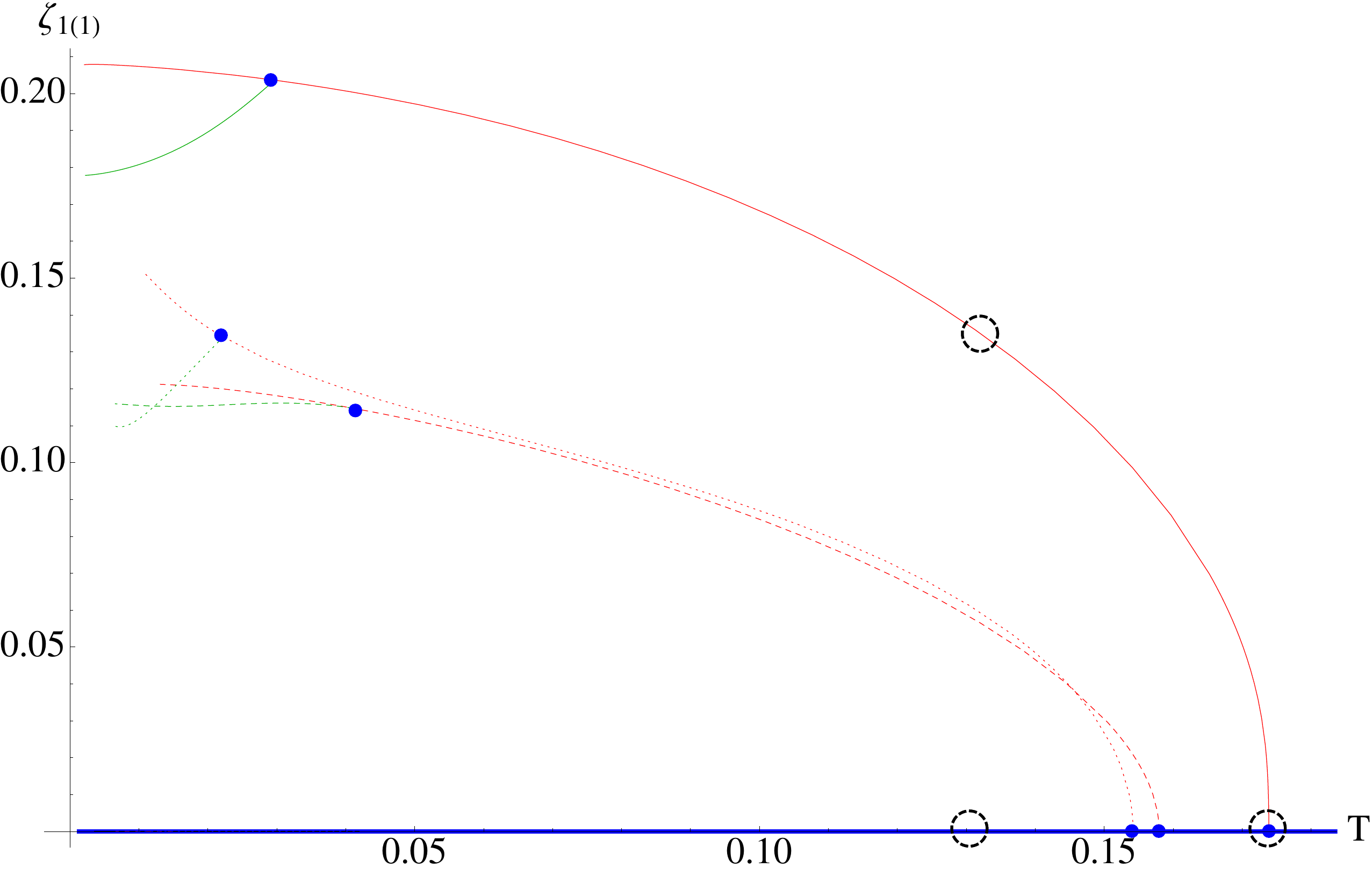}}
\subfloat[A plot of $\zeta_{2(2)}\sim <{\cal O}_{\zeta_2}>$ versus $T$]{\includegraphics[width=8cm]{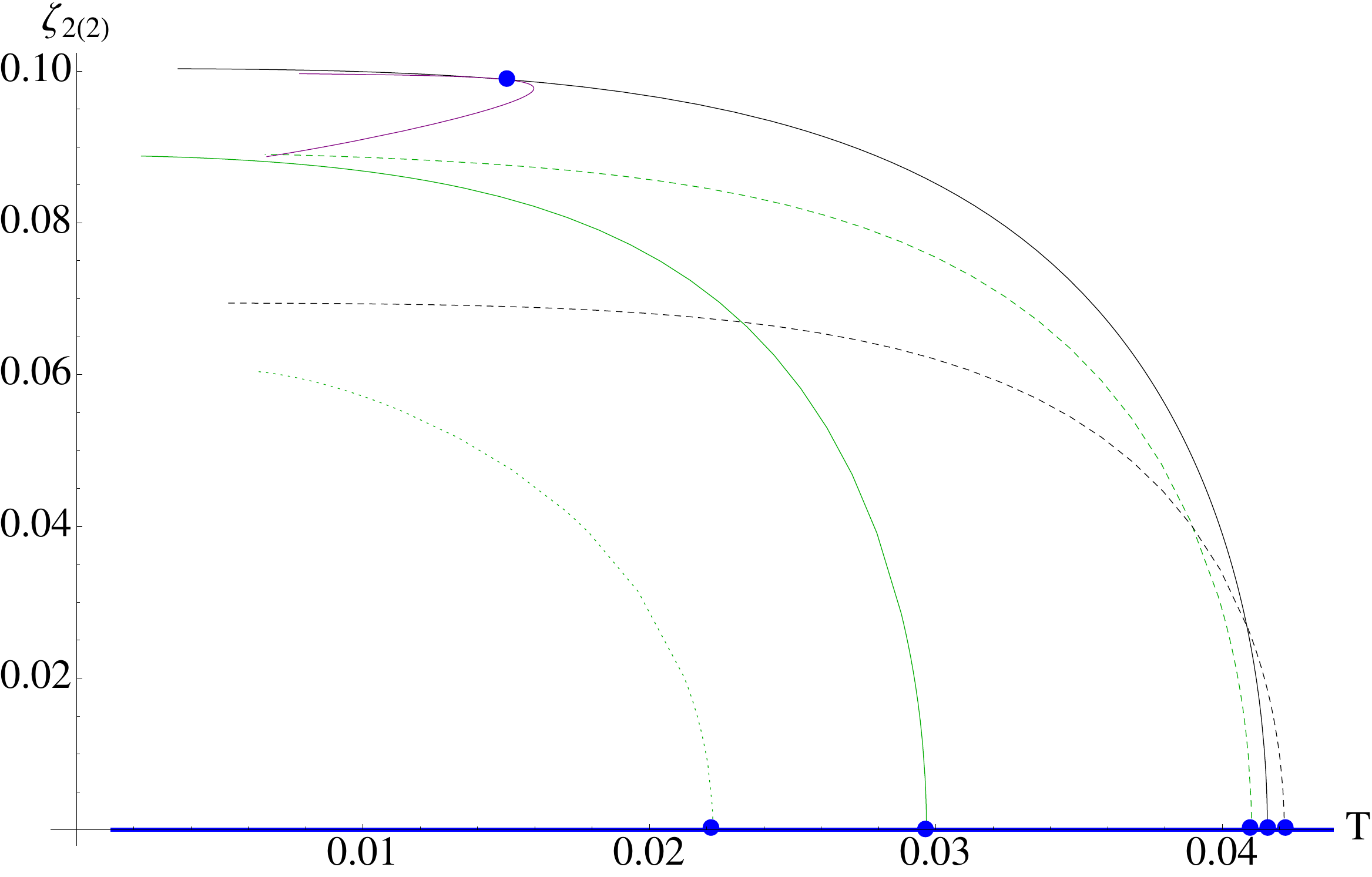}}
\caption{Various black brane solutions in the $SU(3)$ truncation \eqref{su3lag} with $\mu_1=1$, $\mu_0=0$.
The solid red branch is the superfluid black brane associated with an
instability of the AdS-RN black brane and has $\zeta_1\ne 0$ and $\zeta_2= 0$. 
It branches into the solid green branch which also has $\zeta_2\ne 0$. 
The blue dotted and dashed lines are two branches of black branes with $\zeta_1=\zeta_2=0$ 
associated with the Gubser-Mitra instability of the AdS-RN black brane. 
The solid blue line in each of the bottom figures represents
the AdS-RN black branes, but in the top figure it also represents the back reacted Gubser-Mitra 
branes. The solid black line are the superfluid black branes with $\zeta_1=0$ and $\zeta_2\ne 0$ found in
\cite{Gauntlett:2009dn}. Additional black branes are also shown.
The solid dots indicate the points at which a new branch of black branes is appearing on lines that meet.
The dashed circles indicate the
thermodynamically preferred phase transitions within this truncation: as one lowers $T$
the system moves from the AdS-RN branch to the solid red superfluid
branch, then, discontinuously, to the dotted Gubser-Mitra branch.
}\label{su3fig}
\end{figure}

\subsection{Comment}
In this section we have constructed superfluid black branes for four different truncations of $SO(8)$ gauged supergravity,
aiming to track the superfluid instability at $T_c\approx 0.174$ at which the $\Delta=1$ scalars with $U(1)_R$ charge 1 become unstable. We
have found that the result depends greatly on which truncation one uses. We will make some additional comments in the
discussion section, but here we note that we can compare the free energy of all the black branes that we constructed in all of the
truncations.
We find that the overall preferred phase transitions are those which we saw in the $SU(3)$ truncation, namely 
AdS-RN to solid red line superfluid 
branch to dotted Gubser-Mitra branch in figure \ref{su3fig}.

\section{Superfluid black branes with emergent superconformal symmetry at $T=0$}\label{sec5}
The focus of this section will be the construction of superfluid black brane solutions whose zero temperature limit
are smooth domain walls which interpolate between the $SO(8)$ invariant $AdS_4$ vacuum 
and the supersymmetric $SU(3)\times U(1)$ invariant $AdS_4$ vacuum  that we discussed in \eqref{adsfps}.
We will construct these solutions using the $SU(3)$ invariant consistent truncation of \cite{Bobev:2010ib}; more specifically the
sub-truncation \eqref{su3lag}, and in fact we find that all the relevant solutions have $\zeta_1=0$.
As we will see, unlike in previous sections,
the superfluid black branes will have non-vanishing chemical potentials with respect to both $U(1)$ gauge fields $A^0, A^1$.
Furthermore, our construction will have different boundary conditions for the scalars than we have so far considered, and
corresponds to an alternative quantisation scheme of the bulk $SO(8)$ invariant $AdS_4$ theory
 \cite{Klebanov:1999tb} and in particular is no longer dual to the maximal supersymmetric $d=3$ SCFT.

We will first construct the uncharged domain walls that we were first found in \cite{Ahn:2000aq} 
(see also \cite{Corrado:2001nv}). 
We then construct the new charged domain
walls followed by the new superfluid black branes that approach the charged domain walls at zero temperature.

\subsection{Neutral and charged domain walls}
To construct the domain wall solutions it is important to identify the modes in the $SU(3)$ 
truncation \eqref{su3lag} which correspond to irrelevant operators in the $SU\left(3\right)\times U\left(1\right)$ invariant 
$AdS_4$ vacuum.
By analysing the linearised fluctuations of the scalars, in the coordinates used in \eqref{coordsforads4},
we identify two such irrelevant modes:
\begin{align}\label{eq:SU3_modes_irr}
\delta{\zeta}_{1}&=c_{1}\, r^{d_{1}}\nn
\delta{\zeta}_{2}&=c_{2}\, r^{d_{2}},\quad
\delta{\lambda}=\frac{1}{12}\,\left(2+\sqrt{3}\right)\left(1+\sqrt{17} \right)\,c_{2}\,r^{d_{2}}\, ,
\end{align}
with $c_1,c_2$ constants and $d_{1}=\frac{1}{2}\left(\sqrt{17}-3 \right)$,
$d_{2}=\frac{1}{2}\left(\sqrt{17}-1 \right)$, corresponding to operators with scaling
dimensions $\Delta_i=3+d_i$ in the $SU(3)\times U(1)$ invariant vacuum  (see \cite{Bobev:2010ib}).
The $SU\left(3\right)\times U\left(1\right)$ invariant vacuum breaks one of the two $U(1)$ symmetries
and so we expect a linear combination of the vector fields will be massive in this vacuum
and therefore give rise to another irrelevant operator. Indeed we find that there is 
a massive vector dual to an operator with scaling dimension $\Delta=2+d_2$ and the
corresponding irrelevant deformation is
\begin{align}\label{gfidef}
\delta{A}^{0}&=c_{3}\,r^{d_{2}},\qquad
\delta{A}^{1}=\frac{\sqrt{3}}{2}\,c_{3}\,r^{d_{2}}\, ,
\end{align}
with $c_3$ constant.

We now recover the neutral domain wall solutions first found in \cite{Ahn:2000aq}.
This flow is a supersymmetric flow and was solved in \cite{Ahn:2000aq} using first order
flow equations utilising a superpotential. Here we will solve the equations of motion directly.
In \eqref{su3lag} we use the ansatz \eqref{eq:rad_ansatz} for the metric, we set $A^0=A^1=0$ and take
all scalars to be functions of $r$ only and $\zeta_i$ real. 
It turns out that to generate a flow with the $SO\left(8\right)$ solution as the UV endpoint we
should shoot out from 
the $SU\left(3\right)\times U\left(1\right)$ $AdS_4$ vacuum using just one of the two modes in \eqref{eq:SU3_modes_irr} 
namely $c_{1}=0, c_2\ne 0$ . Thus we will set $\zeta_1=0$ and 
impose the boundary conditions as $r\to 0$
\begin{align}
{\zeta}_{2}&=\frac{1}{\sqrt{3}}+c_{2}\, r^{d_{2}}+\dots \nn
{\lambda}&=2-\sqrt{3}+\frac{1}{12}\,\left(2+\sqrt{3}\right)\left(1+\sqrt{17} \right)\,c_{2}\,r^{d_{2}} +\dots\nn
g&=\frac{3\sqrt{3}}{4}+\dots\nn
\beta&=\beta_0+\dots\, .\label{expndw}
\end{align}
Using a shooting method we find that for appropriately chosen $(c_2,\beta_0)$ we can indeed hit the $SO(8)$ fixed 
point
and we can then read off the boundary data using the expansion \eqref{eq:SO8_exp4}. 
We find that if in the IR we choose 
 \begin{align}
 c_2\approx -1.41,\qquad\beta_0\approx 0.26\, ,
 \end{align}
 then in the UV we have
 \begin{align}
 {\zeta}_{2(1)}&\approx0.27,\qquad \zeta_{2(2)}\approx-0.08\nn
 \lambda_{(1)}&=\frac{1}{10},\qquad \lambda_{(2)}\approx-0.03\nn
 m&\approx-0.08
 \end{align}
 and we note that, for convenience, we have used a scaling symmetry to set
$\lambda_{(1)}=\frac{1}{10}$. This domain wall solution can be viewed as interpolating between a deformation of the maximally supersymmetric
 $SO(8)$ theory, with deformation parameters fixed by $\zeta_{2(1)}$ and $\lambda_{(2)}$ corresponding to
 $\Delta({\cal O}_{\zeta_2})=2$,  $\Delta({\cal O}_{\lambda})=1$,
to the $N=2$ supersymmetric $SU(3)\times U(1)$ theory. It can also be viewed in alternative
 quantisations.

After constructing the neutral domain wall solution we now investigate the possibility of charged domain walls by 
shooting out from the $SU(3)\times U(1)$ fixed point with $c_3$ also non-zero i.e. also using the 
irrelevant mode \eqref{gfidef}. We are interested in charged domain wall solutions that spontaneously break the abelian
symmetry and hence can arise as a zero temperature limit of superfluid black branes.
We find such solutions with $c_{2}\neq 0,c_{3}\neq 0$ and $c_{1}=0$. Thus
we again set $\zeta_1=0$ and use an expansion as in \eqref{expndw} supplemented with
\begin{align}
{A}^{0}&=c_{3}\,r^{d_{2}}+\dots \nn
{A}^{1}&=\frac{\sqrt{3}}{2}\,c_{3}\,r^{d_{}}+\dots\, .
\end{align}
Using a shooting method, we find that there is a particular solution 
for which we can set $\zeta_{2\left(2\right)}=0$ and moreover we don't find any solutions
with $\zeta_{2\left(1\right)}=0$. Thus, 
we find that there is a charged domain wall
solution corresponding to $<{\cal O}_{\zeta_2}>\ne 0$, spontaneously breaking the $U(1)\times U(1)$ 
symmetry to a single $U(1)$, 
in the alternative quantisation
scheme in which $\Delta({\cal O}_{\zeta_2})=1$ (recall that in the maximally supersymmetric quantisation
that we considered in sections 3 and 4, $\Delta({\cal O}_{\zeta_2})=2$).
 
In more detail we find that if we choose the parameters in the IR to be
\begin{align}
c_{2}\approx-0.76,\qquad
c_{3}\approx1.12,\qquad
\beta_{0}\approx1.45\, ,\label{cdwir}
\end{align}
then the UV parameters are given by
\begin{align}\label{cdwuv}
\zeta_{2(1)}&=0.26,\qquad \zeta_{2(1)}=0,\nn
\lambda_{(1)}&\approx  0.38,\qquad \lambda_{(2)}\approx- 0.20\nn
\mu_{0}&\approx0.98,\qquad q_{0}\approx 0,\nn
\mu_1&=1,\qquad\qquad q_{1}\approx-0.42,\nn
m&\approx-0.74\, ,
\end {align}
where this time we used the scaling symmetry to set $\mu_{1}=1$. It is curious that $q_{0}\approx 0$}.

\subsection{Superfluid black branes that approach the $SU(3)\times U(1)$ $AdS_4$ fixed point in the IR at 
zero temperature}
We would like to construct superfluid black branes that have the charged domain wall solution
that we just constructed, interpolating between the $SO(8)$ $AdS_4$ and the $SU(3)\times U(1)$ $AdS_4$ fixed points, 
as the zero temperature ground state. The first point to notice is that the
asymptotic UV values in \eqref{cdwuv} mean that we should aim to construct black branes with non-vanishing
chemical potentials for both $U(1)$ symmetries 
\begin{align}\label{bdmu}
\mu_{0}=0.98\ldots\qquad \mu_{1}=1\, .
\end{align}
As we have noted, in order to spontaneously break $U(1)\times U(1)\to U(1)$,
we must use the alternative, non-supersymmetric quantisation in which $\Delta({\cal O}_{\zeta_2})=1$.
Now we could either quantise $\lambda$ so that 
$\Delta({\cal O}_{\lambda})=1$ (as in the maximally supersymmetric quantisation) or $\Delta({\cal O}_{\lambda})=2$. 
In fact we will find appropriate superfluid black branes that are thermodynamically favoured 
(over unbroken phase black branes) only in the latter quantisation and so we will impose the following constraint on the associated deformation parameter
\begin{align}\label{bdlam}
\lambda_{(1)}=0.38\ldots\, .
\end{align}
The free energy density in this quantisation scheme is
$w=m$.
Note that the Smarr formula for the supersymmetric quantisation given in \eqref{smarr} is still valid, since it doesn't depend on the quantisation used. This is simply because the Smarr formula is derived by writing the on-shell value of the bulk action in two different
ways.

The first task is to construct unbroken phase black branes, preserving the $U(1)\times U(1)$ symmetry, with 
boundary behaviour consistent with \eqref{bdmu} and \eqref{bdlam}. These solutions can be found by setting
$\zeta_{1}=\zeta_2=0$ and then shooting as usual. In figure \ref{fig:VEVs_AltSU3}, we have plotted some features of these solutions
in blue. In particular, we see from figure \ref{fig:VEVs_AltSU3}(c), which plots the Ricci scalar at the horizon 
versus temperature, that at zero temperature these solutions are becoming nakedly
singular.

Switching on $\zeta_2$ (but still keeping $\zeta_1=0$), further numerical analysis reveals that an additional branch of superfluid
black branes emerges at a critical temp $T_c\approx 0.216$. We have plotted some features of these solutions
in red in figure \ref{fig:VEVs_AltSU3}. As one lowers the temperature the solutions approach the charged domain wall
solution that was constructed in \eqref{cdwir} and \eqref{cdwuv} 
as indicated in figure \ref{fig:VEVs_AltSU3} by dots. We have also calculated the
free energy of both solutions finding that the superfluid black branes are thermodynamically preserved.

We have not found any additional branches of black brane solutions, even when we allow for the possibility that
$\zeta_1\ne 0$. Thus, within the particular quantisation scheme we are using, and within the truncation that we 
considering, we conclude that at zero temperature, the system is described by the charged domain wall and in particular
in the far IR has an emergent superconformal symmetry described by the $SU(3)\times U(1)$ invariant $AdS_4$ fixed point.

\begin{figure}
\centering
\subfloat[A plot of $\lambda_{(2)}\sim <{\cal O}_{\lambda}>$ versus $T$]{\includegraphics[width=7cm]{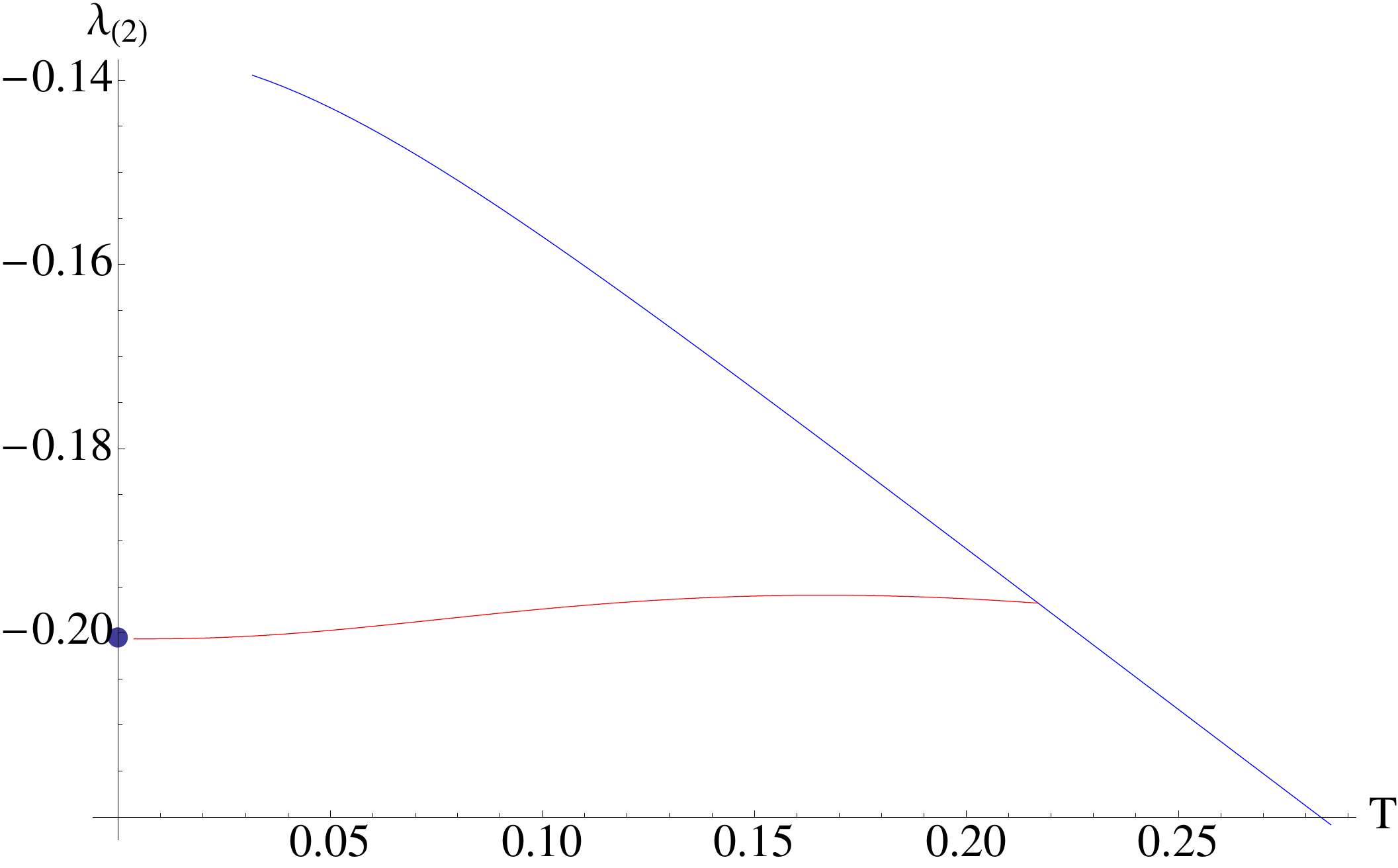}}
\subfloat[A plot of $\zeta_{2(1)}\sim <{\cal O}_{\zeta_2}>$ versus $T$]{\includegraphics[width=7cm]{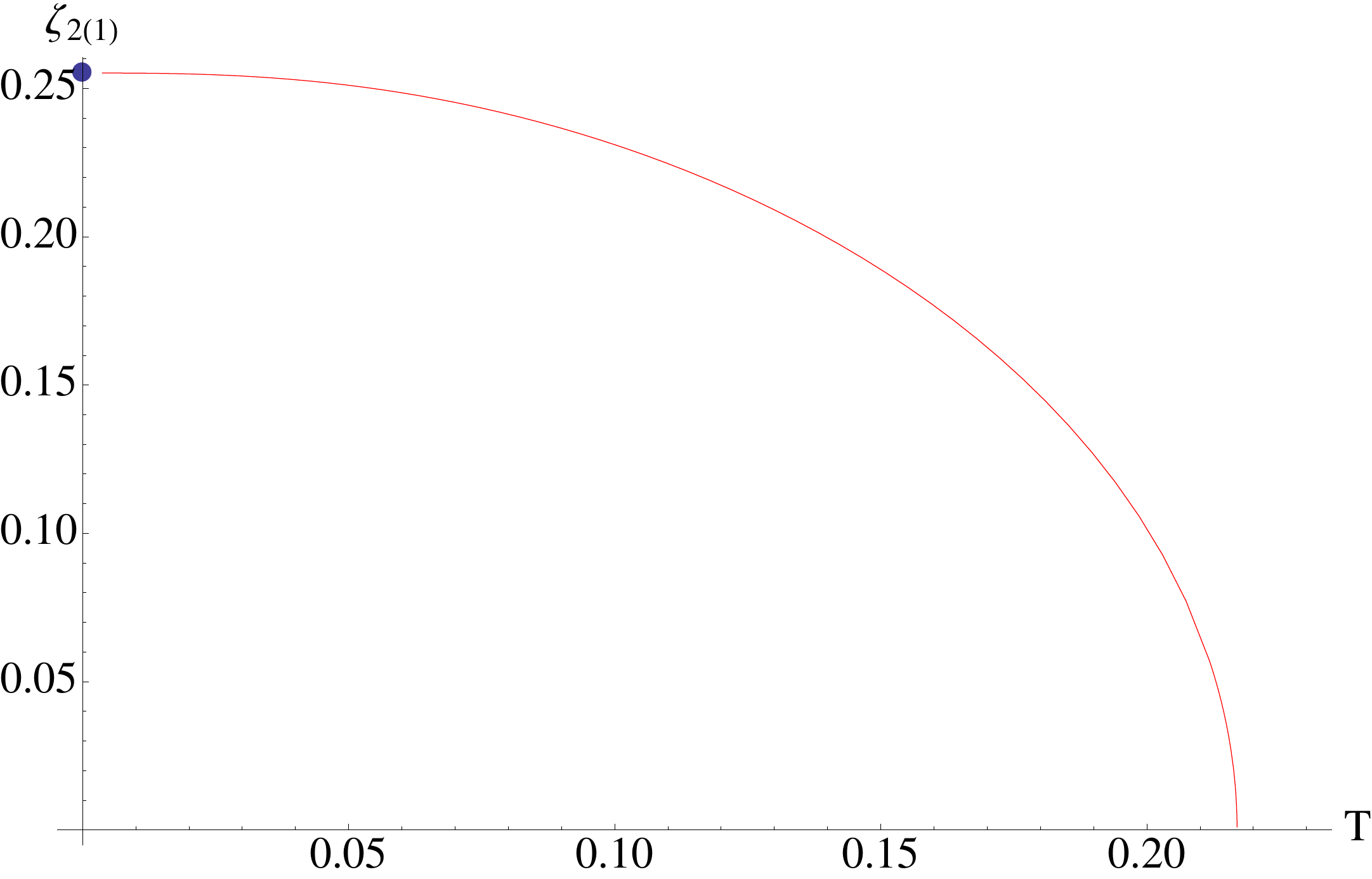}}\\
\subfloat[The free energy density versus $T$]{\includegraphics[width=7cm]{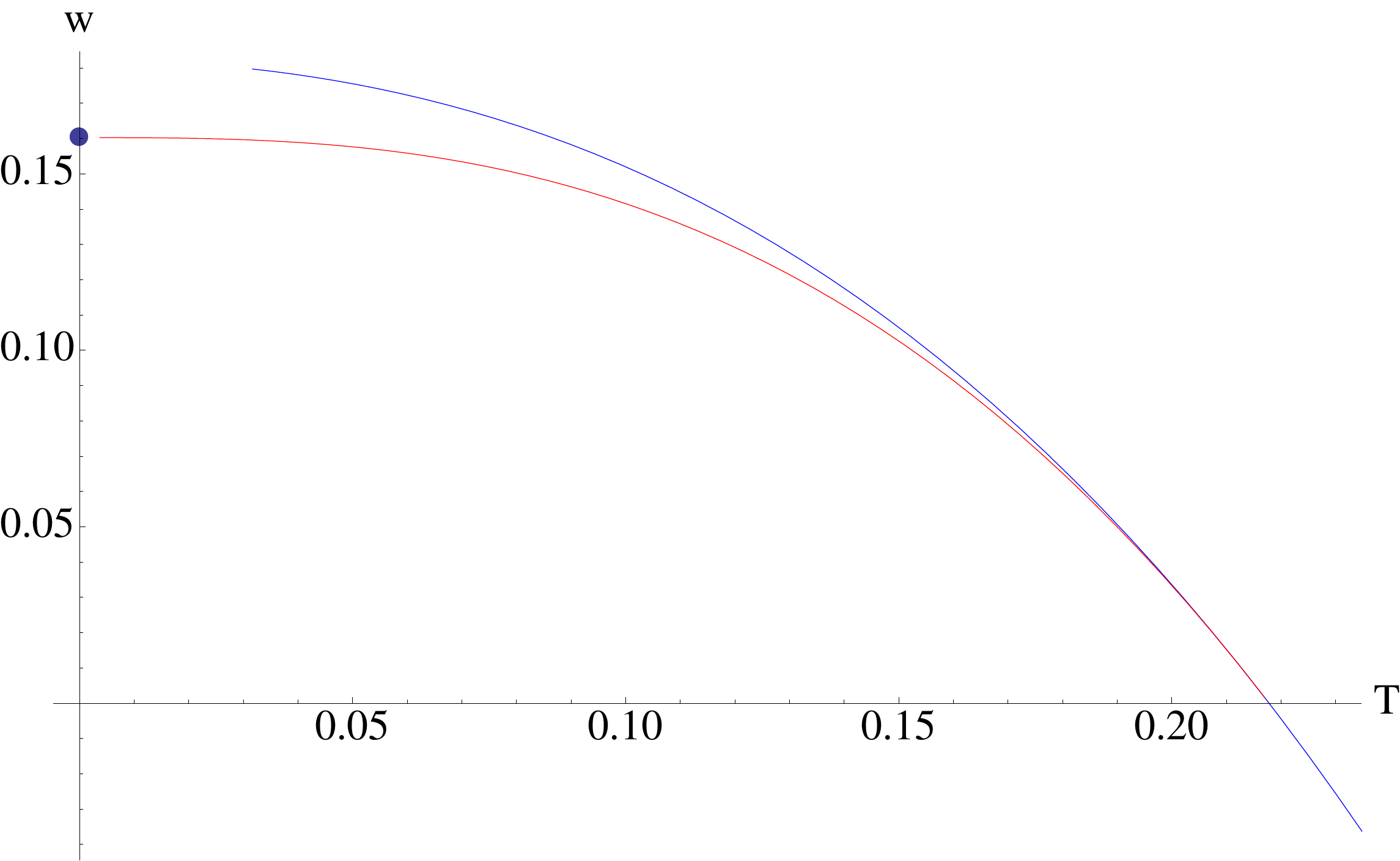}}
\subfloat[A plot of the Ricci scalar evaluated at the horizon of the black brane versus $T$]{\includegraphics[width=7cm]{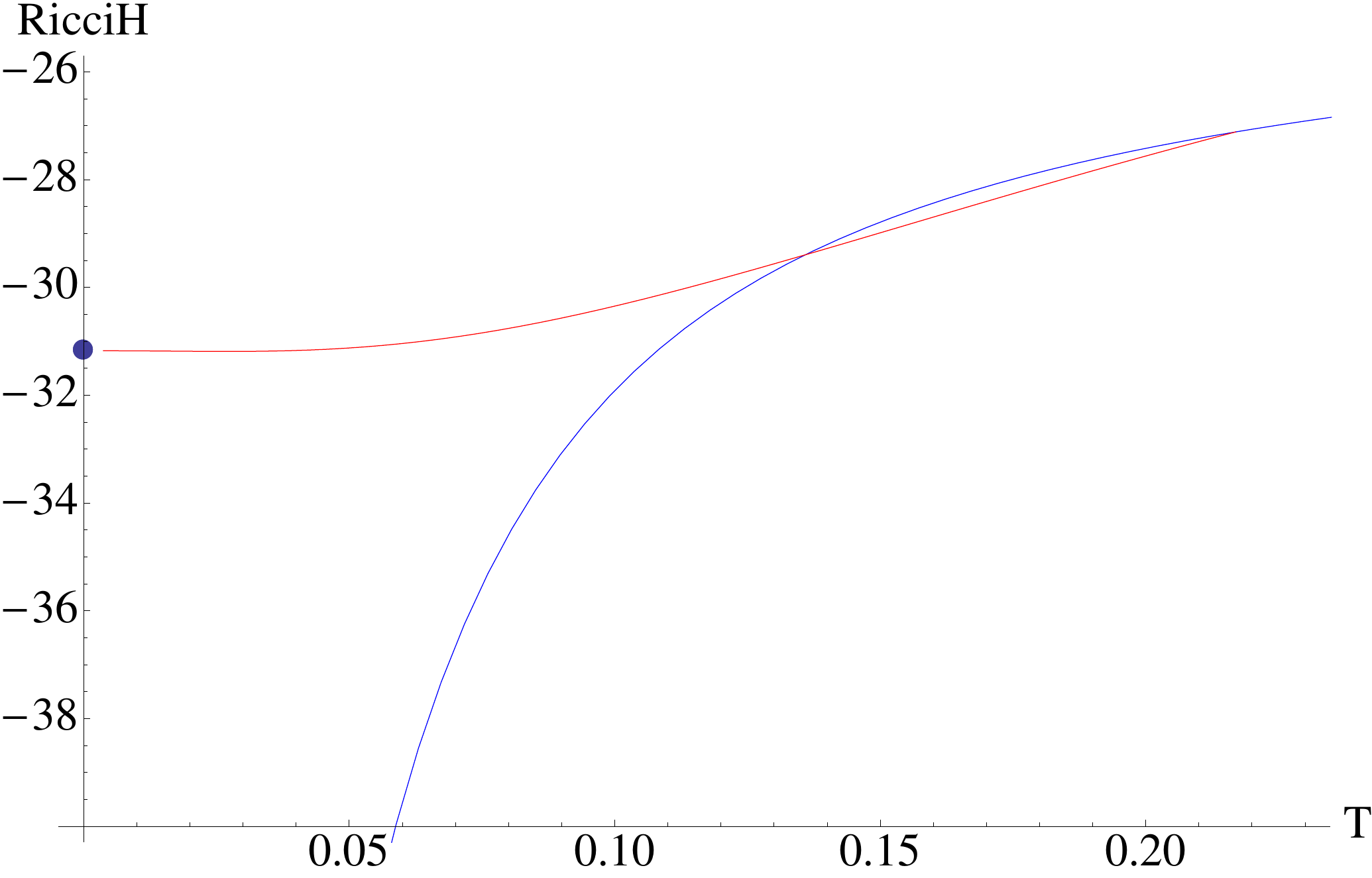}}
\caption{Black brane solutions of the $SU(3)$ invariant truncation \eqref{su3lag} all of which have $\zeta_1=0$. The two chemical potentials are $\mu_{0}\approx0.98, \mu_{1}=1$. We are using an alternative quantisation
in which $\Delta({\cal O}_{\zeta_2})=1$ and $\Delta({\cal O}_{\lambda})=2$ and the deformation parameter $\lambda_{(1)}$ is given by $\lambda_{(1)}\approx0.38$. The blue branch is the unbroken phase black branes and the red branch is the superfluid black branes.
The solid dots at zero temperature are the relevant quantities for the charged domain walls that interpolate between
the $SO(8)$ $AdS_4$ and the $SU(3)\times U(1)$ $AdS_4$ vacua.
}\label{fig:VEVs_AltSU3}
\end{figure}

\section{Discussion}\label{sec6}
In this paper we have constructed a variety of black brane solutions of $D=11$ supergravity
associated with the maximally supersymmetric $d=3$ CFT dual to $AdS_4\times S^7$
when held at finite temperature and non-vanishing chemical potential with respect to 
a Reeb $U(1)_R\subset SO(8)$ 
symmetry. We have found that both the types of black branes which
exist and also, perhaps surprisingly, 
their thermodynamical properties strongly depend upon which truncation of $SO(8)$ gauged
supergravity that one is using.

Consider, for example, the superfluid black branes associated
with the $U(1)_R$ charge 1 and $\Delta=1$ instability which has the highest known critical temperature.
In the 4 equal charged scalars truncation the black branes only 
exist for temperatures greater than the critical temperature 
and are not thermodynamically preferred (see the solid red line in figure \ref{fig:charged_VEVs_single}). 
Yet in the other truncations, they exist for temperatures lower
than the critical temperature and are thermodynamically preferred over the AdS-RN black branes (see the solid red lines
in figures \ref{fig:charged_VEVs_N12} and \ref{su3fig}). 

We have also seen, for the first time in a top down setting, that superfluid black branches can sprout additional
branches of black branes as one lowers the temperature. For example we see in figures \ref{fig:charged_VEVs_N12} and \ref{su3fig} the solid
green branch, which breaks both $U(1)$ symmetries,  emerging from the solid red branch which just breaks one of the
$U(1)$ symmetries.
Again the details strongly depend on the truncation

In addition we studied black branes associated with the Gubser-Mitra instability. 
One interesting point is that in contrast to the minimally coupled
charged scalars of \cite{Denef:2009tp}, the instability does not reveal itself when 
analysing the linearised modes in the near horizon $AdS_2$ region of the
AdS-RN black branes at zero temperature. We also constructed the corresponding back reacted black brane solutions and
again we saw markedly different behaviour depending on the truncation (see the dotted and dashed blue lines on figures 
\ref{fig:charged_VEVs_N12} and \ref{su3fig}). We also saw that these branches can then grow a variety of different superfluid branches.

In addition to the expected second order phase transitions we have shown
that first order transitions are also possible. For example,
in the $SU(3)$ truncation, which contains the thermodynamically preferred black branes of
all the truncations that we considered, the system jumps from
the superfluid branch to the Gubser-Mitra branch (solid red line to doted blue line in figure \ref{su3fig}). This example underscores the point that in principle one may need to construct all of the black brane solutions, and not just those continuously connected to the instability at highest temperature, in order to map out the full phase diagram.

The solutions that we have found in the four different truncations are, together, rather complicated and it is not clear
how they are all related. It is possible that a simpler and more unified
picture will emerge if one is able to study the problem directly in the context of $SO(8)$ gauged supergravity,
and is able to exploit the full symmetry of the system. The overall thermodynamically preferred solutions that
we constructed become nakedly singular in the IR (the Gubser-Mitra dotted blue branch in figure \ref{su3fig}).
It would be interesting to know whether the full $SO(8)$ gauged supergravity has additional black brane
solutions that have non-singular zero temperature limits.

Of course it may be that modes outside of $SO(8)$ gauge supergravity play an important role
and hence the relevant black branes would need to be constructed directly in $D=11$ supergravity.
If that is the case then the direct approach we have undertaken may not be tractable, and
alternative techniques and approaches will need to be developed to establish, for example, the
ultimate zero temperature ground states. Perhaps
it is possible to develop the kinds of ideas discussed in \cite{Douglas:2009zn}, 
for example.
It also might be the case that the example of $AdS_4\times S^7$ is a particularly
difficult one to examine, and it may be that one can make more progress by studying CFTs dual to
other Sasaki-Einstein manifolds. The homogeneous examples of $AdS_4\times Q^{1,1,1}$ \cite{D'Auria:1983vy} or 
$AdS_4\times M^{3,2}$ 
\cite{Witten:1981me}\cite{Castellani:1983mf} might be good examples to study further, for example.

We have also constructed the first top down black brane solutions that at zero temperature approach a charged
domain wall solution which approaches a supersymmetric $AdS_4$ vacuum in the IR. More specifically, the
charged domain walls interpolate between the $SO(8)$ invariant $AdS_4$ vacuum in the UV in an 
alternative quantisation scheme and the supersymmetric
$SU(3)\times U(1)$ $AdS_4$ vacuum in the IR.
It would be interesting to know whether or not this is the true ground state within this quantisation scheme 
both within $SO(8)$ gauged supergravity and more generally.

\subsection*{Acknowledgements}
We would like to thank Nikolay Bobev, Michael Douglas, 
Johanna Erdmenger, Eric Perlmutter, Jorge Russo, Julian Sonner, Daniel Waldram and Toby Wiseman
for helpful discussions. 
AD is supported by an EPSRC Postdoctoral Fellowship.
JPG is supported by an EPSRC Senior Fellowship and a Royal Society Wolfson Award. 
JPG would like to thank the Simons Center for Geometry
and Physics for hospitality, where some of this work was done.

\appendix

\section{Black brane thermodynamics for the $SU(3)$ invariant truncation}\label{secapp}

Here we will discuss some relevant thermodynamical formula for the $SU(3)$ invariant truncation
\eqref{su3lag}, following the presentation given\footnote{In comparing with \cite{Gauntlett:2009dn} one should note that the definitions of
the function $g$ slightly differ.} in \cite{Gauntlett:2009dn,Gauntlett:2009bh}. 
We will consider the radial and purely electric configurations discussed at the beginning of section 4.4, and
in particular the metric is given by \eqref{eq:rad_ansatz}. It will be convenient to reinstate $\beta_\infty$ in the asymptotic
expansion \eqref{eq:SO8_exp4}, which just adds $\beta_\infty$ to the expansion of $\beta$ and in addition
we have $a_1=e^{-\beta_\infty}(\mu_1+q_1/r+..)$ and similarly for $a_0$.
We can show that the on-shell bulk euclidean action reads
\begin{equation}
I_{OS}=\Delta\tau\, \mathrm{vol_{2}}\,\int_{r_{+}}^{\infty}dr\,\left[2r^{3}ge^{-\beta/2} \right]^{\prime}\, ,
\end{equation}
where $\tau=it$ and $\mathrm{vol_{2}}=\int dx dy$. The temperature of the black brane is 
given by $T=\frac{r_{+}^{2}}{2\pi}\,g^{\prime}\left(r_{+}\right)\,e^{\left(\beta_{\infty}-\beta\left(r_{+}\right) \right)/2}$ and we have
$\Delta\tau=e^{\beta_{\infty}/2}/T$.

The counter term that we need to add in order to obtain finite action, and to ensure that
$\Delta({\cal O}_{\zeta_1})=\Delta({\cal O}_{\lambda})=1$ and $\Delta({\cal O}_{\zeta_2})=2$,
reads
\begin{equation}
I_{ct}= \int_{\partial M}d\tau d^{2}x\sqrt{g_{\infty}}\,\left[-K+2\sqrt{2}+\sqrt{2}\,\left(-\zeta_{1}^{2}+\zeta_{2}^{2}-3\lambda^{2}-\sqrt{2}\,n^{\mu}\left(\lambda\partial_{\mu}\lambda+\zeta_{1}\partial_{\mu}\zeta_{1} \right) \right) \right]
\end{equation}
where $g_\infty= \lim_{r\to \infty} \sqrt{2}r^3 e^{-\beta/2} g^{1/2}$ and
$K=\lim_{r\to\infty}g^{\mu\nu}\nabla_\mu n_\nu$ is the trace of the extrinsic curvature. For our ansatz 
this gives
\begin{align}
I_{ct}=& \Delta\tau\, \mathrm{vol_{2}}\lim_{r\rightarrow\infty}\,e^{-\beta/2}\Bigg[-r^{2}e^{\beta} \left( r^{2}ge^{-\beta} \right)^{\prime} -4r^{3}g+4r^{3}g^{1/2}\nn
&+2r^{3}g^{1/2}\left(-\zeta_1^{2}+\zeta_2^{2}-3\lambda^{2}  - 2rg^{1/2} \left(\zeta_1\zeta_1^{\prime}+3\lambda\lambda^{\prime} \right) \right)\Bigg]\, .
\end{align}
Plugging in the expansion \eqref{eq:SO8_exp} (after reinstating $\beta_\infty$) we find the value of the total on-shell action reads
\begin{align}\label{afour}
I_{tot}=I_{OS}+I_{ct}=
& \frac{\mathrm{vol_{2}}}{T}\,\left[m-4\,\zeta_{2(1)}\zeta_{2(2)}  \right]\, .
\end{align}

In order to derive a Smarr type of formula we write the bulk Euclidean action in the equivalent form
\begin{align}
I_{OS}=\Delta\tau\, \mathrm{vol_{2}}\,\int_{r_{+}}^{\infty}dr\,\left[r^{2}e^{-\beta/2}\left(\left( r^{2}g\right)^{\prime} -r^{2}g\beta^{\prime}\right)  +  r^{2}e^{\beta/2}C^{\alpha\gamma}a_{\alpha}^{\prime}a_{\gamma}      \right]^{\prime}
\end{align}
where
\begin{align}
C^{00}=&\frac{1-2\lambda+6\lambda^{2}-2\lambda^{3}+\lambda^{4}}{\left(\lambda-1\right)\left(1+\lambda \right)^{3}}\nn
C^{11}=&\frac{1+2\lambda+6\lambda^{2}+2\lambda^{3}+\lambda^{4}}{\left(\lambda-1\right)\left(1+\lambda \right)^{3}}\nn
C^{01}=C^{10}=&-2\sqrt{3}\lambda\frac{1+\lambda^{2}}{\left(\lambda-1\right)\left(1+\lambda \right)^{3}}\, .
\end{align}
Plugging in the expansion \eqref{eq:SO8_exp4} (after reinstating $\beta_\infty$) we obtain the alternative formula
\begin{equation}\label{eq:os_action_2}
I_{tot}=\frac{\mathrm{vol}_{2}}{T}\,\left( -2m-2\pi r_{+}^{2}T+q_{0}\mu_{0}+q_{1}\mu_{1}+24\lambda_{1}\lambda_{2}+8\zeta_{1(1)}\zeta_{1(2)}+4\zeta_{2(1)}\zeta_{2(2)}\right)\, .
\end{equation}
Comparing the two equivalent expressions \eqref{eq:os_action_1} and \eqref{eq:os_action_2} we conclude that
\begin{align}\label{smarr}
-3m-sT+\mu_{0}q_{0}+\mu_{1}q_{1}+24\lambda_{1}\lambda_{2}+8\zeta_{1(1)}\zeta_{1(2)}+8\zeta_{2(1)}\zeta_{2(2)}=0\, ,
\end{align}
where we defined the entropy density
$s=2\pi r_{+}^{2}$.

We next consider the variation of the total action obtaining equations of motion plus boundary terms. For the on-shell
variation we obtain 
\begin{align}\label{eq:os_action_1}
\delta I_{tot}
=&\frac{\mathrm{vol}_{2}}{T}\,\left(m-\frac{1}{2}\mu_{0}q_{0}-\frac{1}{2}\mu_{1}q_{1}
-12\lambda_{(1)}\lambda_{(2)}-4\zeta_{1(1)}\zeta_{1(2)}-2\zeta_{2(1)}\zeta_{2(2)} \right)\,\delta\beta_{\infty}\notag \\
&+\frac{\mathrm{vol}_{2}}{T}\,\left(q_{0}\delta\mu_{0}+q_{1}\delta\mu_{1}+12\lambda_{(1)}\delta\lambda_{(2)}+4\zeta_{1(1)}
\delta\zeta_{1(2)}-4\zeta_{2(2)}\delta\zeta_{2(1)} \right)\, .
\end{align}
In this variation we are keeping $\Delta\tau$ fixed and so $\delta\beta_\infty=2\delta T/T$. We now see that the total action 
is stationary for fixed $T, \mu_0,\mu_1$ and for fixed deformation parameters $\lambda_{(2)}, \zeta_{1(2)}, \zeta_{2(1)}$.

Finally, if we define the free energy via $W=T[I_{tot}]_{OS}=w\mathrm{vol_2}$, we have 
$w=w(T,\mu_0,\mu_1,\lambda_{(2)},\zeta_{1(2)}, \zeta_{2(1)})$. Using \eqref{afour}, \eqref{smarr} and \eqref{eq:os_action_1}
we obtain
\begin{align}
\delta w=-s\delta T+q_{0}\delta\mu_{0}+q_{1}\delta\mu_{1}+12\lambda_{(1)}\delta\lambda_{(2)}+4\zeta_{1(1)}
\delta\zeta_{1(2)}-4\zeta_{2(2)}\delta\zeta_{2(1)}\, .
\end{align}


\bibliographystyle{utphys}
\bibliography{sfluidrefs}{}
\end{document}